\documentclass[conference]{IEEEtran}
\IEEEoverridecommandlockouts
\usepackage{cite}
\usepackage{amsmath,amssymb,amsfonts}
\usepackage{algorithmic}
\usepackage{graphicx}
\usepackage{textcomp}
\usepackage{xcolor}

\usepackage{hyperref}
\usepackage{booktabs}
\usepackage{listings}
\usepackage{float}

\usepackage{tikz}
\usetikzlibrary{shapes.geometric, shapes.symbols, arrows.meta, positioning, fit, calc, shadows, backgrounds}

\def\BibTeX{{\rm B\kern-.05em{\sc i\kern-.025em b}\kern-.08em
    T\kern-.1667em\lower.7ex\hbox{E}\kern-.125emX}}
\begin{document}

\title{MemTrust: A Zero-Trust Architecture for Unified AI Memory System\\
}

\author{\IEEEauthorblockN{Xing Zhou}
\IEEEauthorblockA{\textit{Independent Researcher}}
\IEEEauthorblockA{\textit{zhou.xing@live.com}}
\and
\IEEEauthorblockN{Dmitrii Ustiugov}
\IEEEauthorblockA{\textit{NTU Singapore}}
\IEEEauthorblockA{\textit{dmitrii.ustiugov@ntu.edu.sg}}
\and
\IEEEauthorblockN{Haoxin Shang}
\IEEEauthorblockA{\textit{Supermem AI Inc.}}
\IEEEauthorblockA{\textit{hx@supermem.io}}
\and
\IEEEauthorblockN{Kisson Lin}
\IEEEauthorblockA{\textit{Supermem AI Inc.}}
\IEEEauthorblockA{\textit{k@supermem.io}}
}

\maketitle

\begin{abstract}
AI memory systems are evolving toward unified context layers that enable efficient cross-agent collaboration and multi-tool workflows, facilitating better accumulation of personal data and learning of user preferences. However, centralization creates a trust crisis where users must entrust cloud providers with sensitive digital memory data.

We identify a core tension between personalization demands and data sovereignty: centralized memory systems enable efficient cross-agent collaboration but expose users' sensitive data to cloud provider risks, while private deployments provide security but limit collaboration.

To resolve this tension, we aim to achieve local-equivalent security while enabling superior maintenance efficiency and collaborative capabilities. We propose a five-layer architecture abstracting common functional components of AI memory systems: Storage, Extraction, Learning, Retrieval, and Governance. By applying TEE protection to each layer, we establish a trustworthy framework. Based on this, we design MemTrust, a hardware-backed zero-trust architecture that provides cryptographic guarantees across all layers.

Our contributions include the five-layer abstraction, "Context from MemTrust" protocol for cross-application sharing, side-channel hardened retrieval with obfuscated access patterns, and comprehensive security analysis. The architecture enables third-party developers to port existing systems with acceptable development costs, achieving system-wide trustworthiness.
We believe that AI memory plays a crucial role in enhancing the efficiency and collaboration of agents and AI tools. AI memory will become the foundational infrastructure for AI agents, and MemTrust serves as a universal trusted framework for AI memory systems, with the goal of becoming the infrastructure of memory infrastructure.

Evaluation shows MemTrust achieves less than 20\% performance overhead on enterprise workloads while providing local-equivalent confidentiality, enabling context centralization without sacrificing data sovereignty.
\end{abstract}


\section{Introduction}

\subsection{Motivation: The Context Centralization Imperative}

Users have consistently demanded that AI memory data remains secure, with memory storage systems being private or at least trustworthy and fully controllable—particularly for enterprise users. This requirement stems from the sensitive nature of context data, which often contains proprietary information, personal preferences, and business-critical knowledge. Enterprise users, in particular, require ironclad guarantees that their AI memory systems cannot be accessed by unauthorized parties, including cloud providers, third-party vendors, or malicious insiders.

However, the industry's push toward context centralization—aggregating diverse context sources into unified memory systems—amplifies these security demands exponentially. As AI systems evolve to consolidate user interactions, behavioral patterns, and multi-modal observations from fragmented sources (email, documents, instant messaging, databases, and institutional knowledge), the security requirements become even more stringent for enterprise users. This consolidation creates higher-value attack targets, where a single breach could expose comprehensive user profiles and organizational intelligence.

The proliferation of large language models (LLMs) has driven widespread adoption of AI agents in enterprise environments. However, a fundamental tension has emerged between two competing forces shaping the future of AI systems:

\textbf{The Personalization Imperative}: Industry practice shows that when AI systems accumulate and retain rich personal context, they create high switching costs and durable advantage. Long-context memory and personalization increasingly matter more than marginal model capability gains. ChatGPT's memory function exemplifies this trend \cite{OpenAI-Memory-2024}—users develop deep dependency not because of model capability alone, but because the system "remembers who they are." In enterprise scenarios, solutions that leverage proprietary or sensitive data to build context-centered moats are favored because they address domains that general-purpose models cannot easily replicate.

\textbf{The Architectural Shift: Context-Application Decoupling}: The AI industry is converging on an architecture where application containers separate from context/memory. Practitioners anticipate a future in which unified context—spanning preferences, behaviors, history, and multimodal observations—becomes a shared layer that third-party applications can call on-demand, analogous to single-sign-on (e.g., Google/Facebook login) but for context, so new apps reuse existing understanding instead of rebuilding it.

This emerging architecture promises transformative benefits:
\begin{itemize}
\item \textbf{Cross-application personalization}: A unified context layer enables AI agents to understand "who the user is, what they're doing, and what they prefer" regardless of the application front-end.
\item \textbf{Reduced cold-start problems}: New applications can immediately leverage accumulated user context rather than starting from zero.
\item \textbf{Enhanced user experience}: Users interact with a coherent "AI assistant" that follows them across devices and services, rather than isolated, context-less chatbots.
\item \textbf{Cross-tool/provider collaboration}: Applications and AI tools themselves generate new memories; a unified context layer lets heterogeneous tools and providers collaborate seamlessly and handle more complex tasks. For instance, a requirement discussed in the ChatGPT client can be clarified in the Claude client without re-supplying background, because the shared memory carries user profile and history across vendors. Similarly, a design-to-build chain can span Figma for visual decisions, Lovable for product drafts, and Cursor or Claude Code for implementation, with all specs, decisions, and constraints written back to a unified memory. These examples fit the goal of cross-vendor portability rather than a single provider's walled garden.
\end{itemize}

\textbf{The Data Fragmentation Problem}: Despite this personalization imperative, enterprise data remains inherently fragmented across heterogeneous systems: documents reside in SharePoint or Confluence, communications scatter across email (Outlook, Gmail) and instant messaging (Slack, Teams), structured data lives in SQL and NoSQL databases, and institutional knowledge accumulates in wikis and ticket systems. This fragmentation produces multiple pathologies: duplicated and inconsistent representations that create information silos and conflicting responses; inefficient retrieval that requires multiple API calls, driving latency and cost; persistent context loss where traditional LLMs fail to maintain preferences, history, or domain knowledge across conversations, devices, or applications; and a multi-modal gap, as most solutions remain text-dialog centric and lack manageable support for visual, action, and scene contexts—future context must span text, visual memories (images/videos), behavioral telemetry, spatial positioning, and social relationships.

\textbf{The Security Paradox}

This architectural shift toward unified context creates a fundamental security paradox: \textbf{centralizing rich personal context amplifies both its utility and its vulnerability}. The more comprehensive the context, the higher the value to both users (personalization quality) and adversaries (intelligence gathering, profiling, surveillance).

This paradox manifests in a critical deployment dilemma:

\textbf{Cloud deployment}: It offers scalability, cross-device accessibility, and low operational overhead, but it requires trusting cloud infrastructure providers (AWS, Azure, Google Cloud) and AI memory service providers with plaintext access to intimate personal data. However, this trust assumption is increasingly untenable because: (1) strict regulations (GDPR \cite{GDPR-2018} erasure, HIPAA \cite{HIPAA-Privacy-Rule} PHI, China's PIPL) drive deletion and residency obligations; (2) insider threats and data-mining incentives persist (malicious admins, competitive analytics); (3) state-level access mandates remain (US CLOUD Act \cite{US-CLOUD-Act-2018}, China's Data Security Law, other lawful intercepts); and (4) repeated high-profile breaches show "secure cloud storage" is still unresolved.

\textbf{On-premises deployment}: It provides full data control but carries substantial costs: high capital expenditure for hardware (\$15K-50K per server) and data center build-out; recurring operational overhead for IT staff ($\sim$\$500/month part-time for SMEs), security patching, and disaster recovery; and nontrivial power/cooling ($\sim$\$100/month per server). It also breaks the unified-context promise—users struggle to access context seamlessly across mobile devices, third-party apps, or collaborative tools without complex VPNs or exposing local infrastructure to the internet.

Industry validation confirms this tension as a primary barrier. Recent surveys indicate the market increasingly favors AI applications that leverage proprietary or privacy-sensitive data to build competitive moats—targeting scenarios that general-purpose model providers cannot easily address. Successful examples include Harvey (legal), Mercor (recruiting), and Abridge (healthcare), where context is both highly valuable and tightly regulated.

The fundamental question is: \textbf{Can we achieve cloud-native convenience with security guarantees equivalent to local deployment, without requiring trust in cloud providers or AI memory service operators?}

\subsection{AI Agent and Memory System Overview}
\label{sec:system_overview}

In the evolving landscape of autonomous AI, agents are increasingly specialized—optimized for specific modalities such as conversation, coding, or data analysis. To enable these specialized agents to collaborate effectively and provide a cohesive user experience, a unified memory infrastructure is essential. This section outlines the interaction model between the User, multiple AI Agents, and the centralized AI Memory System.

\subsubsection{Cross-Agent Context Sharing}

The core value of the AI Memory System lies in its ability to decouple context from the execution runtime of individual agents. By externalizing state into a secure, unified memory layer, disparate agents can share a common understanding of the user's intent, preferences, and project history. This architecture transforms the AI Memory System into a "Context Hub," allowing a user to start a task with one agent and seamlessly continue it with another without restating information.

\subsubsection{Illustrative Scenario: Collaborative Software Development}

To demonstrate this interaction flow, we consider a software development scenario involving two distinct agents:
\begin{enumerate}
    \item \textbf{Chat Agent:} A conversational interface (e.g., specialized in requirements gathering and brainstorming).
    \item \textbf{Coding Agent:} An engineering interface (e.g., specialized in code generation and refactoring).
\end{enumerate}

\begin{figure*}[htbp]
    \centering
    \begin{tikzpicture}[
        node distance=2.0cm and 3.5cm,
        entity/.style={
            rectangle,
            draw=black!70,
            thick,
            rounded corners,
            minimum width=2.5cm,
            minimum height=1.2cm,
            align=center,
            fill=white,
            drop shadow
        },
        user_node/.style={
            entity,
            fill=gray!10,
            circle,
            minimum width=1.5cm
        },
        memory_node/.style={
            entity,
            fill=blue!10,
            minimum height=3.5cm,
            minimum width=3cm
        },
        agent_node/.style={
            entity,
            fill=green!10
        },
        interaction/.style={
            -{Latex[length=3mm, width=2mm]},
            thick,
            draw=black!70,
            font=\footnotesize\sffamily
        },
        step_label/.style={
            midway,
            fill=white,
            inner sep=1.5pt,
            align=center,
            font=\scriptsize\sffamily,
            text=black!80
        },
        tee_box/.style={
            rectangle,
            draw=green!70!black,
            fill=green!5,
            thick,
            rounded corners,
            minimum width=4.5cm,
            minimum height=4.5cm,
            align=center,
            drop shadow
        }
    ]


    \node[user_node] (user) {\textbf{User}};

    \node[agent_node, above right=0.5cm and 3.5cm of user] (chat_agent) {\textbf{Chat Agent}\\(Requirements)};
    \node[agent_node, below right=0.5cm and 3.5cm of user] (code_agent) {\textbf{Coding Agent}\\(Implementation)};

    \node[tee_box, right=4.0cm of chat_agent, anchor=north west, yshift=1.0cm] (tee_boundary) {};
    \node[above=0.1cm of tee_boundary.north, font=\bfseries\sffamily, text=green!70!black] {TEE Protected};
    \node[memory_node, at=(tee_boundary.center)] (memory) {};
    \node[below=0.2cm of memory.north, font=\bfseries] {Memory System};
    \node[font=\scriptsize, align=center] at (memory.center) {Unified Context\\(Vector + Graph)};


    \draw[interaction] (user) -- node[step_label, sloped, above] {\textbf{1. Prompt}\\ "Idea: Snake Game\\in Python"} (chat_agent.west);

    \draw[interaction, dashed] (chat_agent.east) -- node[step_label, above] {\textbf{2. Store Context}\\ "Topic: Snake Game"\\ "Lang: Python"} ($(memory.west)+(0, 1.0)$);

    \draw[interaction] (user) -- node[step_label, sloped, below] {\textbf{3. Prompt}\\ "Write code for\\game loop"} (code_agent.west);

    \draw[interaction, dashed] ($(code_agent.east)+(0, 0.2)$) -- node[step_label, sloped, above] {\textbf{4. Query}\\ "Current Project?"} ($(memory.west)+(0, -1.0)$);

    \draw[interaction, dashed] ($(memory.west)+(0, -1.4)$) -- node[step_label, sloped, below] {\textbf{5. Retrieve}\\ "Context: Python, Snake"} ($(code_agent.east)+(0, -0.2)$);

    \draw[interaction] (code_agent.south west) to[out=200, in=300] node[step_label, below] {\textbf{6. Response}\\ "Here is the Python code..."} (user.south east);


    \draw[dotted, thick, gray] ($(user.east)+(1.5, 3)$) -- ($(user.east)+(1.5, -3)$) node[below] {Front-end Interface};
    \draw[dotted, thick, gray] ($(chat_agent.east)+(2.0, 1.5)$) -- ($(chat_agent.east)+(2.0, -4.65)$) node[below] {Backend Infrastructure};

    \end{tikzpicture}
    \caption{Multi-Agent Collaboration workflow via TEE-protected Unified AI Memory. The User defines requirements with the Chat Agent (1), which securely persists the context within the TEE boundary (2). Later, the Coding Agent retrieves this shared context (4-5) to fulfill a generic request (3), returning personalized code (6) without redundant prompting, all while maintaining end-to-end confidentiality.}
    \label{fig:multi_agent_interaction}
\end{figure*}
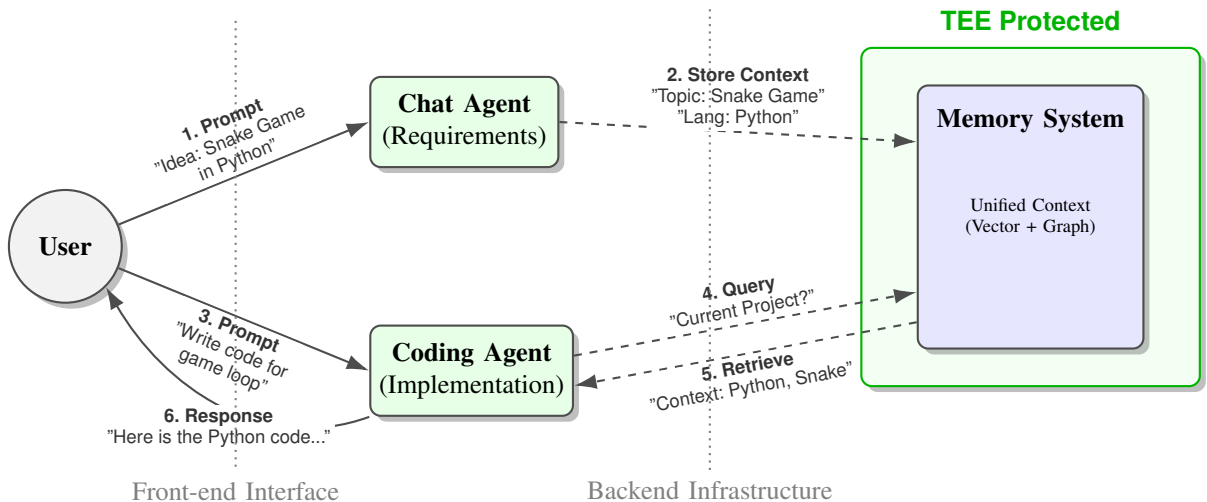

The workflow, depicted in Fig. \ref{fig:multi_agent_interaction}, proceeds as follows:

\begin{itemize}
    \item \textbf{Phase 1: Intent Definition (User $\leftrightarrow$ Chat Agent).}
    The user initiates a session with the \textit{Chat Agent} to brainstorm a new project, specifically a "Python-based Snake Game." The Chat Agent processes this interaction and securely synchronizes the context to the TEE-protected \textit{AI Memory System}. It stores semantic facts (e.g., "Project: Snake Game", "Language: Python", "UI Library: Pygame") into the unified memory with end-to-end confidentiality.

    \item \textbf{Phase 2: Context Handoff (User $\rightarrow$ Coding Agent).}
    The user then switches to the \textit{Coding Agent} to begin implementation. Crucially, the user issues a concise prompt: "Generate the main game loop." The user does not need to specify the programming language or the game type again.

    \item \textbf{Phase 3: Context Retrieval (Coding Agent $\leftrightarrow$ Memory System).}
    Upon receiving the vague instruction, the \textit{Coding Agent} queries the \textit{AI Memory System}. It retrieves the relevant "Project Context" established during the chat session. The memory system returns the structured knowledge that the user is building a "Snake Game in Python."

    \item \textbf{Phase 4: Execution.}
    Armed with this retrieved context, the \textit{Coding Agent} generates the correct Python/Pygame code and returns it to the user, completing the cross-agent workflow.
\end{itemize}

\subsection{Our Approach: Hardware-Backed Zero Trust}

We answer this question affirmatively by leveraging Trusted Execution Environments (TEEs)—hardware-backed security mechanisms that protect data in use through memory encryption and isolated execution. Our system, MemTrust, makes the following core design choices:

\textbf{Zero-trust architecture}: We adopt the principle that cloud infrastructure providers, service operators, and even privileged insiders cannot be trusted with plaintext access to context data. All sensitive operations—memory extraction from raw sources, consolidation into embeddings and knowledge graphs, and retrieval for agent queries—occur within TEEs. Cryptographic attestation provides verifiable proof of system integrity, enabling users to remotely verify the trustworthiness of cloud-deployed services through remote attestation.

\textbf{Multi-TEE support}: Rather than depending on a single TEE technology, we design abstractions that work across Intel SGX (process-level enclaves), AMD SEV-SNP and Intel TDX (VM-level isolation), AWS Nitro Enclaves (cloud-native enclaves), and ARM CCA (emerging mobile/edge platform). This future-proofs the architecture against vendor lock-in and regional availability constraints.

\textbf{Secure memory lifecycle}: We protect the entire AI memory pipeline:
\begin{itemize}
\item \textbf{Extraction}: LLM-based fact extraction, entity recognition, and embedding generation occur entirely within TEEs
\item \textbf{Storage}: Encrypted memory data is persisted to untrusted backends (vector databases, knowledge graphs) with keys never leaving TEE or user control
\item \textbf{Retrieval}: Query processing, access control enforcement, and result aggregation happen within TEE boundaries
\end{itemize}

\textbf{Cryptographic key management}: Master keys remain under user control via Hold-Your-Own-Key (HYOK) models, or are derived within TEEs using attestation-bound protocols. Data encryption keys are hierarchically derived and sealed using hardware keys, enabling secure persistence across TEE restarts without compromising security.

\textbf{Enabling the Decoupled Context Vision}

Critically, our architecture directly enables the industry's vision of context-application decoupling:

\textbf{Unified context across applications}: Different AI applications (writing assistants, code generators, shopping agents) can access the same TEE-protected context layer, subject to fine-grained access control policies enforced within the TEE.

\textbf{Privacy-preserving context sharing}: Just as OAuth allows "Login with Google" without Google seeing the third-party application's data, MemTrust enables "Context from MemTrust" without the memory service provider seeing plaintext context. Applications receive attestation-bound tokens that grant access to specific context subsets, with all access logged for audit.

\textbf{Cross-device continuity}: Because context is encrypted at rest and in transit, with keys managed via user-controlled or TEE-sealed mechanisms, users can securely access their unified context from any device—smartphone, laptop, smart home devices—without per-device encryption key management. Device-specific sub-keys are derived from the user master key, and TEE verifies device attestation before releasing context.

The resulting architecture achieves three critical properties:

\begin{enumerate}
\item \textbf{Confidentiality}: Even with root access to cloud infrastructure, adversaries (including cloud providers) cannot decrypt context data or observe computation inside TEEs.

\item \textbf{Integrity}: Remote attestation allows users to verify that their data is processed by genuine, uncompromised TEE implementations running expected code versions—a form of "zero-trust compute" that doesn't require trusting operators.

\item \textbf{Availability}: Despite security protections, the system maintains cloud-native scalability, cross-device accessibility, and acceptable performance overhead (less than 20\% in our evaluation).
\end{enumerate}

\subsection{Contributions}

This paper makes the following contributions:

\textbf{Industry alignment}: We provide the first security infrastructure specifically designed to enable the industry's emerging vision of context-application decoupling. By demonstrating how "context as infrastructure" can operate in zero-trust environments, we bridge the gap between architectural vision and practical deployment (Section 4.6).

\textbf{Threat model}: We formalize the "untrusted cloud provider" threat model for unified AI memory systems, identifying attack vectors ranging from hypervisor snooping to legal data access mandates, and deriving security requirements that balance protection with usability (Section 3).

\textbf{System design}: We present MemTrust, a comprehensive architecture integrating multiple TEE technologies (SGX, SEV-SNP, TDX, Nitro, CCA) with AI memory systems. We describe secure protocols for memory extraction, updating, and retrieval, along with cryptographic key management (HKDF-based hierarchical keys, hardware sealing) and multi-tenant isolation strategies (Section 4).

\textbf{Context sharing protocols}: We design OAuth-like "Context from MemTrust" protocols that enable cross-application context access with fine-grained permissions, attestation-bound tokens, and comprehensive audit logging. This directly addresses the industry need for decoupled context architectures (Section 4.6).

\textbf{Implementation}: We implement a working prototype using AMD SEV-SNP (primary), Intel SGX, and AWS Nitro Enclaves, integrating with production systems: Qdrant (vector DB), SurrealDB (knowledge graph), and LLMs (Gemini 2.5 Flash API proxying, local Qwen 3 8B). The codebase ($\sim$17K lines Rust + Python) demonstrates practical feasibility (Section 6).

\textbf{Evaluation}: We conduct security analysis including attestation verification, penetration testing, and cryptographic audits. Performance evaluation on enterprise workloads (10K documents, 50K emails, 1M knowledge triples) demonstrates <20\% overhead for SEV-SNP compared to non-TEE baselines, with near-linear horizontal scaling (Section 7).

\textbf{Open challenges}: We identify future research directions including GPU-TEE integration for large model inference, formal verification of security protocols, oblivious RAM for access pattern hiding, multi-modal context at scale, and context markets with data sovereignty guarantees (Section 8).

\textbf{Impact and Vision}

This work represents the first integration of TEEs with unified AI memory systems, filling a critical gap at the intersection of AI capabilities and security infrastructure. We envision MemTrust as foundational for the next generation of AI platforms where:

\begin{itemize}
\item \textbf{Personal context becomes a portable, secure asset}—not trapped in walled gardens
\item \textbf{Users control who accesses their context} through cryptographic proof, not legal agreements
\item \textbf{AI personalization quality improves} through richer, multi-source, multi-modal context—without sacrificing privacy
\item \textbf{Enterprise AI adoption accelerates} in regulated industries (healthcare, finance, legal) previously blocked by data sovereignty concerns
\end{itemize}

As the AI industry shifts focus from incremental model gains to accumulating and leveraging context, the security of that context becomes a core competitive differentiator. MemTrust provides the foundation for this transition, enabling unified, intelligent context to coexist with user autonomy and verifiable trust.
We recognize the critical importance of AI memory in improving the efficiency and collaboration of agents and AI tools. As AI memory emerges as the fundamental infrastructure for AI agents, MemTrust provides a universal trusted framework for AI memory systems, aiming to establish itself as the infrastructure of memory infrastructure.

\section{Background and Related Work}

\subsection{AI Memory Systems}

AI memory systems have progressed from research prototypes to commercial platforms, yet rigor and trust are uneven. We summarize evolution, a unifying five-layer view, and security gaps in a related-work style.

\textbf{Evolution across systems:} Early work such as MemGPT \cite{Packer23} and Letta \cite{Letta24} emphasized virtual context and paging to stretch context windows. Engineering-focused systems—Mem0 \cite{Mem0-25}, Graphiti/Zep \cite{Zep-25}, MemMachine \cite{MemMachine-25}, LangMem \cite{LangMem-25}—added multi-backend storage (vector/graph/KV), temporal graphs, and public performance baselines (LOCOMO \cite{LOCOMO-2024}, LongMemEval \cite{LongMemEval-2024}) while largely trusting the operator. A newer wave—MemOS \cite{MemOS-25}, Cognee \cite{Cognee-24}, Mirix \cite{Mirix-25}, and MemU \cite{MemU-25}—treats memory as an OS-level resource, fuses multimodal signals (text/vision/code/interaction), or targets emotional-companion use cases with self-evolving memories stored as "self-growing" folders and lightweight graphs, claiming high retrieval accuracy and low latency. Research-stage systems (MemGAS \cite{MemGAS-25}, MemoryOS \cite{MemoryOS-25}) explore multi-granularity associations and layered persistence but have not demonstrated production-grade security.

\textbf{Five-layer architecture abstraction:} Despite heterogeneity, we model existing AI memory systems as the following five-layer architecture:

\begin{enumerate}
\item \textbf{Storage}: Provides persistent storage for different memory representations, supporting various backend technologies to handle structured and unstructured data efficiently.

\item \textbf{Extraction \& Update}: Processes raw inputs to extract meaningful memory units and manages their integration into the knowledge base, including conflict resolution and consistency maintenance.

\item \textbf{Learning \& Evolution}: Enables dynamic adaptation of memory through continuous learning from user interactions, enabling personalization, pattern recognition, and temporal evolution of knowledge.

\item \textbf{Retrieval}: Implements query processing and memory access mechanisms to fetch relevant information based on user requests, supporting various search strategies and ranking algorithms.

\item \textbf{Governance}: Manages access control, auditing, and compliance requirements to ensure secure and accountable memory operations across different users and contexts.
\end{enumerate}

We can map different existing AI memory systems to the five layers: MemGPT and Letta implement file-based archival storage with self-editing functions for extraction and update, autonomous memory management decisions for learning and evolution, and context paging for retrieval; Mem0 employs hybrid vector/graph/KV storage with a two-phase LLM pipeline for extraction and update, memory consolidation via usage patterns for learning and evolution, and vector-plus-graph ranking for retrieval; Zep and its Graphiti component utilize temporal knowledge graph storage with bi-temporal fact extraction and conflict resolution, temporal evolution and community detection for learning and evolution, and time-aware graph queries for retrieval; Cognee integrates relational, vector, graph, and ontology storage with ECL pipeline and ontology consistency for extraction and update, semantic relationship evolution for learning and evolution, and multi-modal vector-plus-ontology reasoning for retrieval; while MemOS provides multi-modal storage (text/graph/KV/parametric) with type-aware memory extraction, dynamic type conversion and memory lifecycle management for learning and evolution, and unified vector-plus-graph queries for retrieval. This comprehensive mapping demonstrates that our five-layer abstraction effectively captures the functional essence of diverse AI memory systems, with variations primarily manifesting in implementation maturity and feature emphasis across different layers.

\textbf{Security and trust gaps.} AI memory systems face critical security and compliance requirements across all five layers of our architecture, with data leakage representing the primary security risk alongside privacy protection, multi-tenant isolation, verifiable audit trails, and resistance to adversarial attacks; however, current implementations exhibit systematic gaps: while the first four layers show varied maturity with functional implementations, the Governance layer remains particularly weak, and even the other layers lack robust security: embeddings are stored in plaintext or reversible forms enabling reconstruction attacks, data leaks occur through plaintext interactions with LLMs during extraction, retrieval, and learning \& evolution processes, logs lack tamper evidence for forensic analysis, multi-tenant isolation relies on weak namespace-based separation risking data leakage between tenants, compliance claims rest on provider assertions rather than verifiable proofs (with Zep's SOC2 II being the rare exception), and mechanisms for GDPR-compliant data forgetting and key revocation for data localization are absent. These gaps manifest differently across our five-layer architecture: the Storage layer requires encryption-at-rest with independent key management and third-party verification to prevent data reconstruction; the Extraction \& Update layer needs input sanitization to prevent malicious data injection; the Learning \& Evolution layer demands verifiable learning processes to ensure manipulation resistance; the Retrieval layer requires access pattern protection and result integrity verification; while the Governance layer, currently the weakest link relying on namespaces/ACLs and operator trust, urgently needs neutral third-party verification mechanisms. 
No surveyed system integrates third-party/independent attestation and key management to provide neutral verification and protection against untrusted cloud providers and curious service operators, motivating our zero-trust, hardware-backed design for AI memory.

Despite the functional maturity of existing AI memory systems—from MemGPT's OS-inspired memory management treating LLMs as operating systems with context windows as RAM and external storage as virtual memory, Letta's hierarchical structures with in-context and archival memory, Zep's Graphiti temporal knowledge graphs enabling sophisticated temporal queries, and Mem0's hybrid storage architectures with two-stage LLM pipelines—security and compliance remain critical concerns across three dimensions: first, plaintext interactions with LLMs during memory extraction, retrieval, and evolution processes create substantial PII leakage risks, as sensitive personal information may be inadvertently exposed through LLM responses or cached in model contexts accessible to unauthorized agents; second, unreliable data deletion mechanisms fail to meet compliance requirements like GDPR's "right to be forgotten," with MemGPT and Letta's self-editing operations, Zep's temporal knowledge graphs, and Mem0's hybrid storage backends lacking verifiable deletion guarantees across distributed storage layers; third, insufficient storage isolation guarantees undermine multi-tenant deployments, where MemGPT and Letta rely on basic namespace separation, Zep's graph structures may allow cross-tenant inference through shared entities, and Mem0's proprietary implementations leave tenant isolation mechanisms unverified, while academic systems like HippoRAG and multi-agent architectures similarly lack robust isolation primitives despite their collaborative design goals.

To sum up, current AI memory systems exhibit systematic security deficiencies:

\begin{enumerate}
\item \textbf{PII leakage}: Memory systems may inadvertently expose personal information across sessions or to unauthorized agents.

\item \textbf{Insufficient tenant isolation}: Multi-tenant deployments lack robust isolation mechanisms, risking cross-contamination.

\item \textbf{Opaque memory decisions}: Why agents remember or forget information is typically not explainable or auditable.

\item \textbf{Inconsistent deletion}: GDPR's "right to be forgotten" is not uniformly implemented across memory backends.
\end{enumerate}

These gaps motivate our work on TEE-integrated memory systems with verifiable security guarantees.

\subsection{Trusted Execution Environments}

TEEs provide hardware-enforced isolation that protects code and data from privileged software (operating systems, hypervisors) and physical attacks. The trust root for TEE protections—including data protection, key protection, and execution verification—is embedded in the chip hardware by hardware manufacturers, making it neutral and third-party verified.

Multiple TEE technologies provide hardware-enforced isolation for secure computation: Intel SGX enables application-level isolation through enclaves with Memory Encryption Engine (MEE) for AES-128 encryption, Merkle tree integrity protection, and DCAP-based remote attestation, though limited to 128-512MB EPC; Intel TDX offers VM-level isolation similar to SEV-SNP on 4th-gen Xeon processors; AMD SEV-SNP provides full VM memory encryption with Reverse Map Table (RMP) integrity protection and Versioned Chip Endorsement Key (VCEK) attestation, supporting unmodified application migration; AWS Nitro Enclaves deliver isolated compute environments via custom hypervisor with Nitro-specific PCR attestation (where measurements are signed by the Nitro hypervisor rather than directly by CPU hardware, though the root trust anchors to AWS-controlled hardware); NVIDIA GPU TEE enables confidential computing on H100 GPUs with less than 7\% performance overhead for LLM inference; and ARM CCA introduces "Realms" with dynamic memory management for mobile and edge devices, representing the next generation of confidential computing across diverse hardware platforms.

Recent research on TEE-AI integration has explored several directions: confidential inference demonstrates that NVIDIA H100 GPU confidential computing incurs less than 7\% performance overhead for large language model inference, with overhead dominated by PCIe bus encryption rather than computation itself \cite{Zhu-et-al-24}; confidential RAG implementations like C-FedRAG show full retrieval-augmented generation pipelines executing within SGX enclaves, though memory limitations necessitate careful model partitioning \cite{C-FedRAG-24}; and TEE-assisted federated learning systems leverage SGX for secure gradient aggregation, defending against free-rider and replay attacks \cite{Mo-et-al-24}. However, no prior work has integrated TEEs with comprehensive AI memory systems combining extraction, storage, and retrieval across the complete lifecycle.

\section{MemTrust: A Zero-Trust Architecture for Unified AI Memory System}

\subsection{MemTrust Architecture Design}

Building on the system overview presented in Section 1.2, we now detail the MemTrust architecture that enables the secure, unified AI memory system described earlier. While the previous section illustrated how AI agents collaborate through a shared memory infrastructure, this section focuses on the technical implementation that makes this vision possible in a zero-trust environment.

The MemTrust architecture addresses the fundamental trust gap in existing AI memory systems. Traditional cloud-native stacks assume infrastructure providers are trustworthy, but in an era of zero-trust security models, this assumption is untenable. Memory dumps, cold boot attacks, and privileged escalation by malicious administrators pose significant privacy risks to agent memories.

We propose a novel architecture based on confidential computing, leveraging hardware-level Trusted Execution Environments (TEEs, such as AMD SEV-SNP and Intel TDX) to construct an encrypted, isolated "digital safe" for the entire lifecycle of AI agents. To achieve modular system design, clear security boundaries, and targeted performance optimizations, we rigorously reconstruct MemTrust's technical implementation according to the five-layer model defined in Chapter 2: (i) Storage, (ii) Extraction \& Update, (iii) Learning \& Evolution, (iv) Retrieval, and (v) Governance.

This layered design not only logically decouples complex memory processing workflows but also allows physical implementations to map different TEE optimization techniques to the most appropriate layers. For instance, the bottom storage layer utilizes Rust's memory safety features and HugePages optimization to mitigate I/O overhead of encrypted memory; the middle learning layer leverages NVIDIA H100 TEE's confidential inference capabilities to accelerate memory consolidation; while the top governance layer introduces OIDC protocols bound to remote attestation, ensuring only verified users and code can access this confidential memory.

The following sections will detail the design specifics, technology selections, and security-performance considerations for each layer.

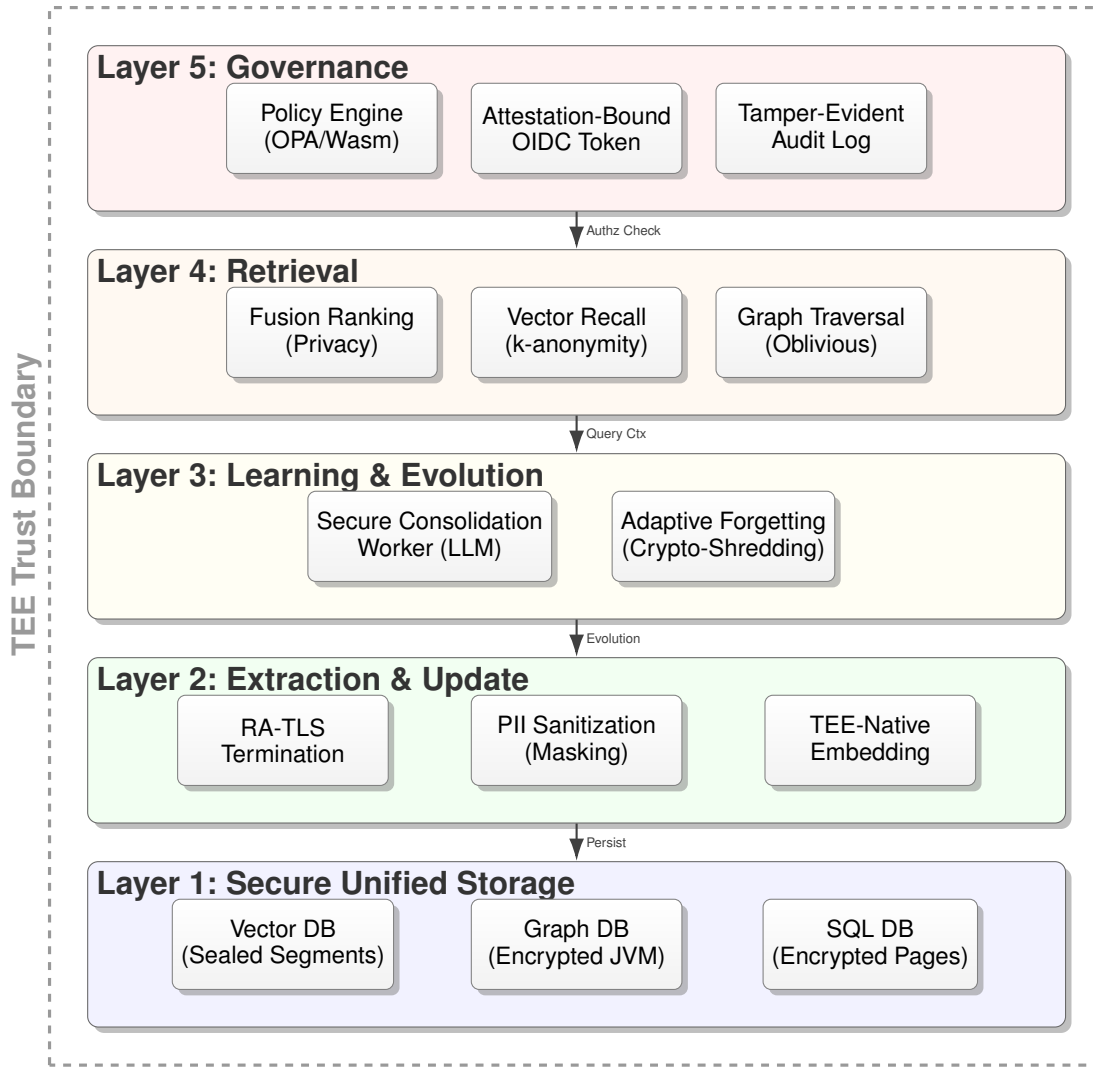
\begin{figure*}[t]
\centering
\begin{tikzpicture}[
    node distance=1.5cm,
    layer/.style={
        rectangle,
        draw=black!60,
        fill=blue!5,
        rounded corners,
        minimum width=13cm,
        minimum height=2.2cm,
        align=center,
        drop shadow,
        inner sep=10pt
    },
    component/.style={
        rectangle,
        draw=black!50,
        top color=white,
        bottom color=gray!10,
        rounded corners=3pt,
        minimum width=2.8cm,
        minimum height=1.2cm,
        font=\small\sffamily,
        align=center,
        drop shadow
    },
    arrow/.style={-{Latex[length=3mm, width=2mm]}, thick, darkgray},
    layerlabel/.style={
        font=\bfseries\sffamily\large,
        anchor=north west,
        text=black!80
    }
]


\node[layer, fill=red!5] (gov) at (0, 10) {};
\node[layerlabel] at (gov.north west) {Layer 5: Governance};
\node[component] (opa) at ($(gov.west)!0.25!(gov.east)$) {Policy Engine\\(OPA/Wasm)};
\node[component] (oidc) at ($(gov.west)!0.5!(gov.east)$) {Attestation-Bound\\OIDC Token};
\node[component] (audit) at ($(gov.west)!0.75!(gov.east)$) {Tamper-Evident\\Audit Log};

\node[layer, fill=orange!5, below=0.5cm of gov] (ret) {};
\node[layerlabel] at (ret.north west) {Layer 4: Retrieval};
\node[component] (rank) at ($(ret.west)!0.25!(ret.east)$) {Fusion Ranking\\(Privacy)};
\node[component] (vec_ret) at ($(ret.west)!0.5!(ret.east)$) {Vector Recall\\(k-anonymity)};
\node[component] (graph_ret) at ($(ret.west)!0.75!(ret.east)$) {Graph Traversal\\(Oblivious)};

\node[layer, fill=yellow!5, below=0.5cm of ret] (learn) {};
\node[layerlabel] at (learn.north west) {Layer 3: Learning \& Evolution};
\node[component] (con) at ($(learn.west)!0.35!(learn.east)$) {Secure Consolidation\\Worker (LLM)};
\node[component] (forget) at ($(learn.west)!0.65!(learn.east)$) {Adaptive Forgetting\\(Crypto-Shredding)};

\node[layer, fill=green!5, below=0.5cm of learn] (ext) {};
\node[layerlabel] at (ext.north west) {Layer 2: Extraction \& Update};
\node[component] (tls) at ($(ext.west)!0.2!(ext.east)$) {RA-TLS\\Termination};
\node[component] (pii) at ($(ext.west)!0.5!(ext.east)$) {PII Sanitization\\(Masking)};
\node[component] (embed) at ($(ext.west)!0.8!(ext.east)$) {TEE-Native\\Embedding};

\node[layer, fill=blue!5, below=0.5cm of ext] (store) {};
\node[layerlabel] at (store.north west) {Layer 1: Secure Unified Storage};
\node[component] (vec_db) at ($(store.west)!0.2!(store.east)$) {Vector DB\\(Sealed Segments)};
\node[component] (graph_db) at ($(store.west)!0.5!(store.east)$) {Graph DB\\(Encrypted JVM)};
\node[component] (sql_db) at ($(store.west)!0.8!(store.east)$) {SQL DB\\(Encrypted Pages)};

\draw[arrow] (gov.south) -- node[right, font=\tiny\sffamily] {Authz Check} (ret.north);
\draw[arrow] (ret.south) -- node[right, font=\tiny\sffamily] {Query Ctx} (learn.north);
\draw[arrow] (learn.south) -- node[right, font=\tiny\sffamily] {Evolution} (ext.north);
\draw[arrow] (ext.south) -- node[right, font=\tiny\sffamily] {Persist} (store.north);

\draw[dashed, very thick, gray!80] ($(store.south west)+(-0.5,-0.5)$) rectangle ($(gov.north east)+(0.5,0.5)$);
\node[rotate=90, font=\bfseries\sffamily\large, text=gray!80] at ($(gov.west)+(-0.8, -5)$) {TEE Trust Boundary};

\end{tikzpicture}
\caption{MemTrust Five-Layer Architecture Overview}
\label{fig:architecture}
\end{figure*}

\subsection{Layer 1: Secure Unified Storage Layer}

The foundational layer of the MemTrust architecture, the Secure Unified Storage Layer, addresses the challenge of persisting heterogeneous memory artifacts—ranging from unstructured documents and structured relational data to high-dimensional vector embeddings and knowledge graphs—while operating under the assumption that the underlying physical storage media and host operating system are untrusted. Unlike traditional encryption-at-rest solutions where keys effectively reside with the cloud provider's managed services, MemTrust mandates that data is always encrypted before leaving the trusted execution boundary and that encryption keys never leave the hardware-protected memory of the TEE. This requirement necessitates a unified storage engine capable of abstracting diverse data modalities into secure cryptographic primitives optimized for the specific I/O constraints of Confidential VMs (CVMs) like AMD SEV-SNP.

To handle unstructured data such as raw documents, images, and conversation logs, MemTrust implements a TEE-native Object Store. Standard object storage access patterns reveal file sizes and access frequencies, which can leak information about user activity. We mitigate this by implementing a chunked, authenticated encryption scheme. Files are segmented into fixed-size blocks (e.g., 4KB or 64KB) and encrypted individually using AES-256-GCM. Crucially, to prevent rollback attacks—where a malicious host presents a valid but outdated version of a file—the system maintains a Merkle Hash Tree (MHT) rooted in the TEE's secure non-volatile memory or a replay-protected monotonic counter service. Every write operation must update the path to the root hash, ensuring that any unauthorized modification or reversion of the data on disk is immediately detected upon the next read. For performance, large binary large objects (BLOBs) are streamed through the TEE using a verified sliding window of hashes, preventing the need to buffer entire large files within the limited encrypted RAM.

For structured relational data, which manages user profiles, session states, and application configurations, MemTrust deploys a hardened SQL engine (customized SQLite or PostgreSQL) running entirely within the secure enclave. While AMD SEV-SNP provides transparent memory encryption, standard database page management can still leak access patterns through the page fault sequence visible to the hypervisor. Our implementation utilizes a user-space encrypted paging mechanism. Database pages are encrypted with a tenant-specific key and a unique page-counter nonce before being flushed to disk. To mitigate the significant I/O overhead associated with frequent encryption operations, the database is configured to use Write-Ahead Logging (WAL) in an append-only mode. This converts random write patterns into sequential writes, which are significantly more performant in TEE environments where the cost of crossing the trust boundary (World Switch) is high.

The most critical component for AI memory, vector storage, presents unique challenges due to the scale of indices required for semantic retrieval. Enterprise vector indices (e.g., HNSW graphs) often exceed the physical capacity of trusted memory. MemTrust implements a "Sealed Segment" architecture for vector databases. The high-dimensional vector space is partitioned into segments. Hot segments, such as the upper layers of an HNSW graph or frequently accessed clusters, are pinned within the TEE's encrypted RAM to maximize search speed. Cold segments are serialized, encrypted, and offloaded to the untrusted host storage. When a query requires traversing into a cold segment, the system utilizes the AMD SEV-SNP VMPL (Virtual Machine Privilege Level) features to securely map and decrypt the segment on-demand. This hierarchical approach balances the latency requirements of real-time retrieval with the security requirement that no raw vector embedding is ever exposed to the host memory bus.

Finally, for knowledge graph storage, which maps complex entity relationships, the primary security risk is topology leakage—where an attacker infers sensitive relationships by observing the structure of the graph (e.g., node degrees). MemTrust addresses this via an oblivious graph storage scheme. Adjacency lists are not stored contiguously based on their actual size but are instead padded to fixed-sized blocks. Dummy edges are inserted to normalize the degree distribution of nodes, and "supernodes" (entities with massive connections) are broken down into indistinguishable chained blocks. This ensures that on the physical disk, a node with two connections looks identical to a node with a hundred connections, effectively masking the social or organizational structure encoded in the graph.

Performance Analysis: The performance overhead of this unified layer is dominated by the "I/O Wall"—the latency introduced when moving data across the isolation boundary between the TEE and the untrusted host. Our benchmarks on AMD EPYC Turin (Zen 5) processors indicate that while the transparent inline memory encryption of SEV-SNP adds negligible latency (approx. 2-3\%), the software-level integrity verification mechanisms (Merkle Trees) and the "bounce buffering" required for disk I/O can reduce throughput by 15-20\% for random access workloads \cite{Li-SEV-SNP-2022}. However, for sequential read/write operations typical of log structuring and large vector scans, the overhead is amortized to under 8\%, making it a viable trade-off for the security guarantees provided.

\subsection{Layer 2: Extraction \& Update}

The Extraction and Update Layer is responsible for the secure ingestion of raw data streams and their transformation into structured memory artifacts. This layer represents the highest risk surface for PII leakage, as it involves processing plaintext user inputs before they are encrypted for storage. In a Zero-Trust architecture, we cannot rely on external API gateways or cloud load balancers for SSL termination, as this would expose the data session to the cloud provider.

To secure the data in transit, MemTrust implements Remote Attestation TLS (RA-TLS). Unlike standard TLS, where the server authenticates with a domain certificate, RA-TLS embeds a cryptographic attestation report into the TLS handshake. This report proves that the server endpoint is a genuine hardware TEE running the specific, unmodified MemTrust codebase. The TLS session is terminated directly inside the enclave memory, ensuring that the raw data stream is never visible to the host network stack or the hypervisor. This establishes a direct, encrypted tunnel between the user's client device and the secure processing core.

Once ingested, the raw text is processed by TEE-Native Extraction Models. However, when these models require LLM inference services, we face the risk of data leakage to external LLM providers. To address this critical security concern, MemTrust provides two alternative solutions:

\textbf{Solution i: NVIDIA GPU TEE with Open-Source Models}  
MemTrust employs a split-execution strategy to minimize PCIe communication overhead in confidential computing. Vector retrieval and context selection occur entirely within CPU TEEs (AMD SEV-SNP) using AVX-512 optimized algorithms, avoiding the need to transfer large vector indices across the encrypted PCIe bus. Only compact context tensors (Top-K retrieved segments) are securely transmitted via TDISP/IDE-encrypted channels \cite{PCI-SIG-TDISP-2023,PCI-SIG-IDE-2022} to NVIDIA GPU TEEs for LLM inference. This approach ensures that model weights and user inputs never leave the trusted execution boundary, while maintaining end-to-end confidentiality for multi-user deployments. Research demonstrates that NVIDIA's confidential computing architecture provides strong isolation guarantees: GPU memory encryption prevents cross-tenant data leakage, and remote attestation verifies that only authorized, unmodified code runs on the GPU \cite{NVIDIA-CC-2024}.

\textbf{Solution ii: TEE-Protected API Proxy with Masking Service}
When external commercial LLM API services must be used, MemTrust implements a TEE-protected API proxy and masking service. This service intercepts and transforms communications between the internal memory system and external LLM APIs through rule-based, text-matching, and TEE-hosted small LLM agent-driven approaches:

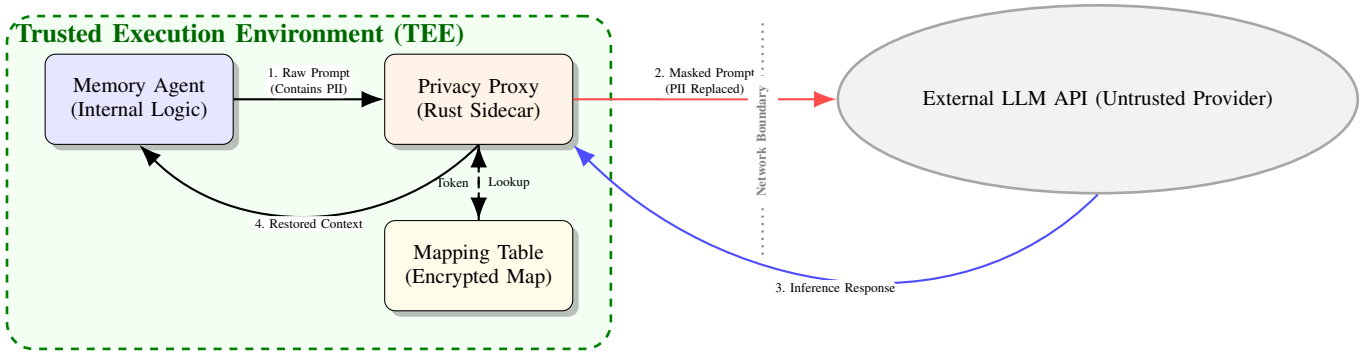
\begin{figure*}[t]
\centering
\begin{tikzpicture}[
    font=\sffamily,
    node distance=1.5cm,
    tee_zone/.style={fill=green!5, draw=green!50!black, dashed,
        line width=1pt, rounded corners=10pt},
    cloud/.style={ellipse, draw=gray!70, fill=gray!10,
        minimum width=4cm, minimum height=2.5cm, line width=1pt},  
    component/.style={rectangle, draw=black, fill=white,
        rounded corners, minimum width=2.5cm, minimum height=1.2cm,
        align=center, drop shadow},
    arrow/.style={-{Latex[length=3mm]}, thick, rounded corners},
    label_text/.style={font=\tiny, align=center, fill=white, inner sep=1pt}
]

\node[component, fill=blue!10, font=\footnotesize] (agent) {Memory Agent\\(Internal Logic)};
\node[component, fill=orange!10, right=2cm of agent, font=\footnotesize] (proxy)
    {Privacy Proxy\\(Rust Sidecar)};
\node[component, fill=yellow!10, below=1cm of proxy, font=\footnotesize] (table)
    {Mapping Table\\(Encrypted Map)};

\node[cloud, right=3.5cm of proxy, font=\footnotesize] (llm) {External LLM API (Untrusted Provider)};

\begin{scope}[on background layer]
    \node[tee_zone, fit=(agent) (proxy) (table), inner sep=0.5cm] (tee) {};
    \node[anchor=north west, text=green!40!black, font=\bfseries]
        at (tee.north west) {Trusted Execution Environment (TEE)};
\end{scope}

\draw[arrow] (agent) -- node[above, label_text]
    {1. Raw Prompt\\(Contains PII)} (proxy);

\draw[arrow, dashed] (proxy) -- node[right, font=\tiny] {Lookup} (table);
\draw[arrow, dashed] (table) -- node[left, font=\tiny] {Token} (proxy);

\draw[arrow, draw=red!70] (proxy) -- node[above, label_text]
    {2. Masked Prompt\\(PII Replaced)} (llm);

\draw[arrow, draw=blue!70] (llm.south) to[out=225, in=315]
    node[below, label_text] {3. Inference Response} (proxy.south east);

\draw[arrow] (proxy.south) to[out=225, in=315]
    node[below, label_text] {4. Restored Context} (agent.south);

\draw[thick, dotted, gray]
    ($(tee.north east)!0.5!(llm.north west)$) --
    ($(tee.south east)!0.5!(llm.south west)$)
    node[midway, fill=white, font=\bfseries\tiny, rotate=90]
    {Network Boundary};

\end{tikzpicture}
\caption{TEE-Protected API Proxy Architecture with PII Masking}
\label{fig:api-proxy}
\end{figure*}

\begin{itemize}
\item \textbf{Outbound Data Masking}: For data sent to external APIs, the service identifies and masks tokens containing PII using pattern matching and lightweight NLP models running within the TEE. Each masked element (e.g., replacing "John Smith" with "[PERSON\_1]") is recorded in an encrypted mapping table that maintains the relationship between masks and original sensitive information.

\item \textbf{Inbound Data Unmasking}: Responses from external LLM APIs are processed within the TEE, where recorded masks are replaced with the original sensitive information before delivery to the internal memory system.
\end{itemize}

While PII masking mechanism described above provides robust protection against direct data exposure, it is important to acknowledge that sophisticated semantic inference techniques could potentially reconstruct contextual information from masked prompts. This hybrid approach is designed as a pragmatic compromise for scenarios requiring advanced LLM capabilities, and should be considered alongside appropriate user consent mechanisms and risk assessments.

This approach provides transparent operation—internal components interact with the proxy as if accessing the external API directly—while preventing PII leakage. The masking service itself runs within the TEE, ensuring that mapping tables and unmasking operations remain confidential.

For high-throughput environments, MemTrust leverages the Confidential Computing capabilities of NVIDIA H100 GPUs to run lightweight extraction models within the TEE boundary. The CPU TEE can offload tensor computations to the GPU TEE through encrypted PCIe channels, protecting model weights and user inputs even during bus traversal.

To prevent traffic analysis attacks—where an observer correlates the timing and size of incoming requests with database updates to infer user activity—the layer incorporates an Oblivious Update Queue. Incoming updates are not applied immediately but are batched into a fixed-rate processing pipeline. If the queue is empty, the system generates dummy update traffic that mimics the statistical profile of real data writes. This decouples the ingress traffic pattern from the storage I/O pattern, blinding external observers to the internal state of the memory system.

\textbf{Performance Analysis}: The computational cost of running extraction models within a CPU TEE (like SEV-SNP) is higher than native execution due to the lack of specialized matrix instructions and the overhead of encrypted memory paging, typically resulting in a 20-30\% inference latency penalty. However, utilizing confidential H100 GPUs drastically reduces this gap, bringing the overhead down to approximately 5-15\%, which is primarily attributed to the PCIe encryption and secure session setup rather than computation. The RA-TLS handshake introduces an initial connection latency of roughly 50-100ms for report verification, but this is a one-time cost per session that does not degrade the throughput of subsequent data streaming.

\subsection{Layer 3: Learning \& Evolution}

The Learning Layer manages the dynamic lifecycle of the AI memory, handling the consolidation of short-term interactions into long-term knowledge and, crucially, the rigorous deletion of data to meet compliance standards. As users interact with the system, memory fragments accumulate and may contain redundancies or contradictions that require synthesis.

MemTrust employs Secure Consolidation Workers that operate as background processes within the TEE. These workers periodically wake up to scan the recent memory logs, decrypting relevant segments into the protected memory space. Using an internal LLM (deployed via the same secure LLM inference approaches described in Layer 2, Solutions i or ii), the system summarizes disparate facts (e.g., updating a user's current project status from multiple meeting notes) and writes the consolidated result back to the storage layer. This entire process occurs without any plaintext data leaving the enclave, protecting the evolution of the user profile from observation.

A unique contribution of our architecture is the implementation of Cryptographic Erasure (Crypto-Shredding) to address the "Right to be Forgotten" (GDPR/CCPA) in an immutable storage system. Physical deletion of data from massive vector indices or distributed logs is technically difficult and nearly impossible to verify in a cloud environment. MemTrust solves this by assigning a unique, granular encryption key (Data Unit Key or DUK) to every distinct memory unit—whether a document, a conversation session, or a specific fact. These keys are managed in a hierarchical Key Vault within the TEE. To "delete" a piece of information, the system simply destroys the specific DUK associated with that data. Without the key, the encrypted data residing on the disk becomes computationally intractable, effectively turning it into random noise. This provides an instantaneous and verifiable deletion mechanism that does not require scrubbing petabytes of storage.

\textbf{Performance Analysis}: The primary cost in this layer is the management overhead of the granular key hierarchy. Storing a unique key for every memory unit increases the metadata storage requirements; for a system with billions of vectors, the key store itself becomes a significant database. To mitigate the latency of fetching keys, we implement a high-performance, oblivious key cache within the TEE. While the background consolidation processes consume CPU cycles, they are scheduled with low priority to avoid impacting user-facing latency. The cryptographic erasure operation is near-instantaneous, offering a massive performance advantage over physical overwrite methods, particularly for large, write-once storage media.

\subsection{Layer 4: Retrieval}

The Retrieval Layer implements a comprehensive index-ranking-retrieval pipeline to fetch relevant context in response to user queries while maintaining strong privacy guarantees. This layer orchestrates multiple recall strategies—keyword-based retrieval, Elasticsearch-powered full-text search, and vector similarity search—within the TEE boundary to ensure end-to-end confidentiality.

The retrieval process follows a three-stage pipeline: \textbf{indexing}, \textbf{ranking}, and \textbf{retrieval}. During indexing, incoming memory artifacts are processed through multiple indexing strategies simultaneously. Keyword indices are built using inverted term-document mappings, Elasticsearch indices leverage full-text analysis with stemming and synonym expansion, vector indices employ HNSW (Hierarchical Navigable Small World) graphs for approximate nearest neighbor search, and graph database indices capture entity relationships and knowledge structures using property graphs or RDF triples with optimized adjacency list representations.

When processing a user query, the system executes parallel multi-path recall within the TEE. \textbf{Keyword recall} retrieves documents containing exact or stemmed query terms from the inverted index. \textbf{Elasticsearch recall} performs full-text search with relevance scoring based on TF-IDF and BM25 algorithms, handling fuzzy matching and phrase queries. \textbf{Vector recall} computes embeddings for the query and performs similarity search in the vector space, retrieving semantically related content even when lexical overlap is minimal. \textbf{Graph database recall} traverses knowledge graphs using graph algorithms such as breadth-first search or shortest path finding, retrieving contextually related entities and relationships that capture complex interdependencies between concepts, enabling multi-hop reasoning and relationship discovery.

To address the critical risk of side-channel attacks through memory access pattern leakage, MemTrust implements a Side-Channel Hardened Multi-Path Retrieval. Each recall strategy is modified to incorporate noise and oblivious processing primitives. For vector search, the HNSW algorithm uses a "Greedy-with-Noise" traversal strategy that probabilistically visits decoy nodes and selects suboptimal paths, injecting entropy into the memory access trace. Vector retrieval operates in fixed-size batches, fetching "buckets" of candidate vectors that include both relevant and dummy entries to obfuscate access patterns against snapshot adversaries.

For keyword, Elasticsearch, and graph database retrieval, we employ oblivious data structures and constant-time operations. Index traversals and graph traversals use conditional moves (CMOV) instead of data-dependent branches to prevent control flow leakage, ensuring that the execution path does not reveal whether specific edges exist or match query criteria. Critical index structures and graph adjacency matrices are cache-pinned within the CPU's L1/L2 cache or maintained in encrypted memory to mitigate cache-timing attacks such as Prime+Probe.

The retrieved candidates from all recall paths are then ranked within the TEE using a fusion algorithm that combines relevance scores from different modalities. This ranking process considers keyword matching strength, semantic similarity, graph-based relationship centrality and path relevance, temporal recency, and user-specific preferences, all computed on decrypted data within the secure boundary.

\textbf{Performance Analysis}: The multi-path recall architecture and side-channel hardening introduce significant computational overhead. The "Greedy-with-Noise" strategy and batched retrieval increase memory bandwidth requirements by 2x to 3x compared to baseline insecure implementations, resulting in proportional reductions in Queries Per Second (QPS). However, for enterprise applications handling sensitive intellectual property or private data, this security-performance tradeoff is often acceptable. In less sensitive deployment scenarios, administrators can tune the noise parameter to balance between maximum security and acceptable performance degradation.

\subsection{Layer 5: Governance}

The final layer, Governance, enforces the root of trust and manages the interface between the MemTrust system and external applications. It ensures that every access is authorized, policy-compliant, and auditable.

We embed a Policy-as-Code Engine (based on Open Policy Agent - OPA) directly within the TEE. Access policies, written in Rego, are hashed and bound to the enclave's measurement. This guarantees that the policies cannot be silently modified by a cloud administrator; any change to the policy file would alter the TEE's measurement, invalidating the remote attestation and causing client connections to fail. This binds the logic of access control to the hardware root of trust.

For accountability, MemTrust maintains a Tamper-Evident Audit Log. Every access request, successful or denied, is logged. These logs are cryptographically chained, where the hash of the current entry depends on the hash of the previous one. Periodically, the "head hash" of this chain is signed by the TEE's private key and anchored to an external immutable ledger, such as a public blockchain or a transparency log (like Trillian). This prevents a "forking attack" where a malicious administrator deletes the most recent logs to hide unauthorized access; the gap between the internal state and the publicly anchored hash would be immediately detectable.

To support the "Context-Application Decoupling" vision, this layer issues Attestation-Bound OIDC Tokens. When an external agent (e.g., a coding assistant) requests access to memory, it receives a token that is cryptographically bound to the specific TLS channel and TEE session. This prevents token theft and replay attacks, as the token cannot be used outside the specific secure connection it was issued for.

\textbf{Performance Analysis}: The governance layer adds minimal latency to the data path. The OPA policy evaluation within the TEE is highly efficient, adding sub-millisecond overhead. The primary cost involves the cryptographic operations for log signing and the initial attestation handshake. Generating a full hardware attestation report on AMD SEV-SNP can take 10-20ms. To avoid incurring this penalty on every request, MemTrust uses a session-based approach: the heavy attestation is performed once during connection establishment to derive a symmetric session key, reducing subsequent per-request governance checks to lightweight symmetric cryptography operations.

\section{Implementation}

To evaluate the feasibility and performance of the proposed architecture, we developed a full-stack prototype of MemTrust. The system serves as a practical instantiation of the zero-trust principles outlined in Section 3, comprised of approximately 20k lines of code—split between a high-performance Rust core (60\%) for cryptographic primitives and storage orchestration, and a Python application layer (40\%) for AI logic and agent interactions. This chapter details the functional baseline of our memory engine—specifically its dual-layer cognition and dynamic adaptation mechanisms—independent of the security layer. We then describe how we mapped this complex, stateful system onto the MemTrust zero-trust infrastructure using AMD SEV-SNP, a hybrid Rust/Python codebase, and cryptographic isolation primitives.

\subsection{Functional Design: The Dual-Layer Cognitive Engine}

Before addressing the security implementation, we first describe the functional capabilities of the underlying memory engine we designed. This engine was built to address the limitations of stateless context windows by mimicking human cognitive processes through a Dual-Layer Cognition architecture. This design bifurcates memory into an Episodic Layer and a Profile Layer, creating a synthesis of precise historical fidelity and high-level semantic abstraction.

The Episodic Layer functions as the system's autobiographical stream, capturing the raw, chronological sequence of user interactions, tool usage, and environmental observations. It does not merely log text but captures the full "state" of an interaction, including the specific AI tools invoked and the immediate outcomes. Conversely, the Profile Layer acts as the semantic distillation of these episodes. Through background consolidation processes, the engine continuously synthesizes raw episodes to extract durable "traits," such as user preferences, professional domain knowledge, and recurring behavioral patterns. This separation allows the system to recall specific past events (e.g., "What code did I write last Tuesday?") while simultaneously applying general knowledge (e.g., "The user prefers Python type hints") to new tasks without needing to re-read the entire history.

To support this cognitive duality, the engine employs a polyglot storage strategy that integrates three distinct database paradigms. A vector database handles high-dimensional embeddings for semantic similarity search, enabling the retrieval of conceptually related memories even when terminology differs. Simultaneously, a graph database maps the structural relationships between entities—connecting users to projects, organizations, and specific documents—which facilitates multi-hop reasoning that vector similarity alone cannot achieve. Supplementing these is a traditional relational database (SQL), which ensures ACID compliance for critical metadata, configuration states, and strict keyword indexing. This tripartite storage ensures that the memory system can handle the multimodal complexity of enterprise data.

\textbf{Episodic Memory (The Stream of Experience)}: This layer captures the raw, time-ordered stream of user-agent interactions. Every user query, tool execution result (e.g., a Python script output or a web search result), and agent response is serialized into a discrete "Episode." Unlike simple chat logs, Episodes are rich objects containing metadata such as timestamps, source applications (e.g., VS Code vs. Slack), and interaction intent. This layer utilizes a Vector Database to index episodes by semantic similarity, enabling the agent to recall "what we did last Tuesday" or "how we solved the SQL error in the previous session" regardless of the time elapsed.

\textbf{Profile Memory (The Semantic Graph)}: While Episodic Memory handles raw experiences, Profile Memory distills these experiences into structured facts and beliefs. We utilize a Graph Database to model the user's world. Nodes represent entities (e.g., "Project Alpha," "John Doe," "Python Preference"), and edges represent relationships with confidence scores (e.g., User --PREFERS--$>$ Concise\_Code). This layer provides the "Static Context" that stabilizes the agent's personality and domain knowledge, preventing hallucinations by grounding generation in a structured knowledge graph.

A critical feature of this engine is its dynamic learning and adaptive forgetting mechanism. Unlike append-only logs that grow indefinitely, our system implements automated relevance evaluation based on retrieval frequency and user feedback. Obsolete or incorrect information—such as changed programming preferences—is logically pruned or archived rather than simply overwritten.

The system operates through an autonomous learning loop: as new Episodes arrive, background "Memory Consolidation" processes extract facts to update the Profile Graph. To prevent memory saturation while maintaining retrieval relevance, we implement an Adaptive Forgetting Mechanism modeled on the Ebbinghaus Forgetting Curve \cite{Ebbinghaus-1885,Murre-2015}. Each memory unit receives a "Memory Strength" (S) that decays over time (t) according to R = e$^{-t/S}$. Unlike biological memory, our system actively reinforces memories—successful retrievals and useful outcomes increment stability S. Data below retrieval thresholds moves to "Cold Storage" (compressed SQL tier), preserving it for forensic recall while freeing active context.

\subsection{TEE-Native System Architecture}

Migrating out memory engine's complex, multi-modal architecture into a Zero-Trust environment requires navigating the "Split-World" constraint of Confidential Computing: the high-level AI logic (often Python-based) requires flexibility, while the security guarantees require low-level control over memory and I/O (typically Rust/C++).

We implemented MemTrust on AMD SEV-SNP (Secure Encrypted Virtualization-Secure Nested Paging) utilizing a layered architecture that leverages AMD's VMPL (Virtual Machine Privilege Level) isolation. Our prototype system is built on AMD EPYC Turin (Zen 5) processors, which provide enhanced AVX-512 support and improved TEE performance characteristics compared to previous generations.

\textbf{Architecture Overview}: Following AMD's SVSM (Secure VM Service Module) pattern, we implement a two-tier privilege separation within the Confidential VM:

\begin{enumerate}
\item \textbf{SVSM-based Sentinel (VMPL0 - Rust)}: The hardware root of trust, implemented as a Secure VM Service Module. This privileged layer handles key management, remote attestation, and secure network termination via RA-TLS. Written in Rust ($\sim$4,500 lines of code), it runs at the highest privilege level (VMPL0) and provides essential security services to the guest environment. The Sentinel is launched during VM initialization and establishes the cryptographic foundation for the entire system.

\item \textbf{Gramine-based Cognitive Engine (VMPL1 - Python/Rust)}: The application layer running as a standard guest environment at VMPL1. We use Gramine Library OS \cite{Gramine-2024} as a lightweight alternative to a full Linux kernel, providing syscall compatibility while minimizing the Trusted Computing Base (TCB). This layer hosts the Vector/Graph databases and Python-based AI orchestration logic. Gramine enables unified deployment across different TEE technologies (SEV-SNP, SGX, TDX), ensuring attestation consistency through its manifest-based measurement system. To mitigate Python fork() overhead in TEE environments where copy-on-write is unsupported, we employ a pre-forked worker pool architecture: Python interpreters are initialized once during system startup and remain resident in VMPL1, with tasks distributed to available workers rather than spawning new processes per request.
\end{enumerate}

\begin{figure*}[t]
\centering
\begin{tikzpicture}[
    node distance=0cm,
    vmpl0/.style={rectangle, draw=red!60!black, top color=red!5, 
        bottom color=red!15, minimum width=10cm, minimum height=2.8cm, 
        align=center, rounded corners},
    vmpl1/.style={rectangle, draw=blue!60!black, top color=blue!5, 
        bottom color=blue!15, minimum width=10cm, minimum height=4.2cm, 
        align=center, rounded corners},
    hardware/.style={rectangle, draw=black!70, fill=gray!20, 
        minimum width=12cm, minimum height=1.5cm, align=center, drop shadow},
    module/.style={rectangle, draw=black!50, fill=white, rounded corners, 
        minimum width=2.2cm, minimum height=1cm, font=\small, drop shadow},
    arrow/.style={->, >=stealth, thick, line width=1.5pt}
]

\node[hardware] (hw) {\textbf{Hardware Layer}\\AMD EPYC (Zen 5) + AMD Secure Processor (PSP)};

\node[vmpl0, above=0.8cm of hw] (v0) {};
\node[anchor=north west, text=red!60!black, font=\bfseries] at (v0.north west) 
    {VMPL0: Sentinel (Root of Trust - Rust)};

\node[module] (keys) at ([xshift=-3cm, yshift=-0.2cm]v0.center) {Key Mgmt (HYOK)};
\node[module] (attest) at ([xshift=0cm, yshift=-0.2cm]v0.center) {Attestation Agent};
\node[module] (net) at ([xshift=3cm, yshift=-0.2cm]v0.center) {RA-TLS Terminator};

\node[vmpl1, above=0.5cm of v0, minimum height=4cm, minimum width=12cm] (v1) {};
\node[anchor=north west, text=blue!60!black, font=\bfseries] at (v1.north west)
    {VMPL1: Cognitive Engine (User Space)};

\node[module, fill=green!10, minimum width=4cm] (py_worker) at ([yshift=0.3cm]v1.center)
    {Python Worker Pool (LlamaIndex Logic)};
\node[module, fill=yellow!10, below left=0.6cm and 0.5cm of py_worker] (qdrant)
    {Qdrant (Vector)};
\node[module, fill=yellow!10, below right=0.6cm and 0.5cm of py_worker] (surreal)
    {SurrealDB (Graph)};

\begin{scope}[on background layer]
    \node[rectangle, draw=black!40, dashed,
          fit=(v1),
          inner sep=0.2cm,
          label={[text=gray, font=\small]above:Gramine Library OS Wrapper}
         ] (gramine_box) {};
\end{scope}

\draw[arrow, dashed, blue!70!black] (v1.south) -- 
    node[right, font=\tiny, align=left] {Shared Memory /\\VirtIO Vsock} (v0.north);

\draw[arrow, red!70!black] (v0.south) -- 
    node[right, font=\tiny, align=left] {RMP Check /\\VCEK Signing} (hw.north);

\node[left=0.5cm of v0, font=\bfseries, text=red!60!black, rotate=90] {Privileged};
\node[left=0.5cm of v1, font=\bfseries, text=blue!60!black, rotate=90] {Restricted};

\end{tikzpicture}
\caption{VMPL0/1 Architecture}
\label{fig:vmpl}
\end{figure*}
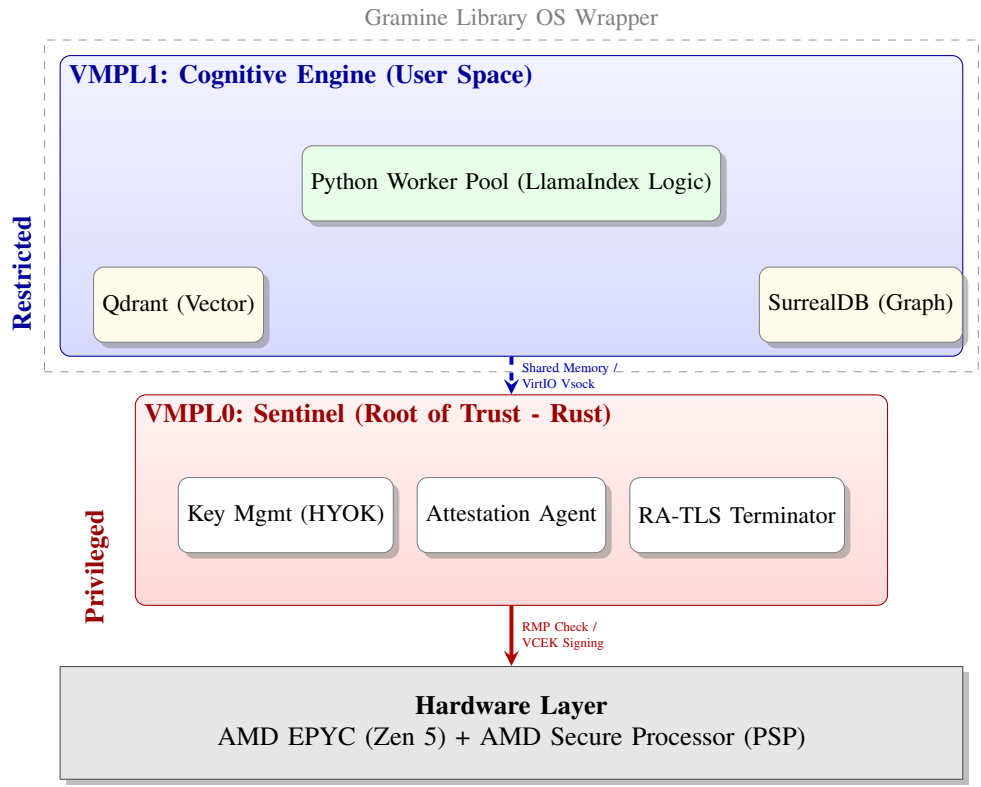

The following subsections detail the implementation of the 5-layer architecture within this protected enclave.

\subsubsection{Securing the Storage Layer (Layer 1)}

The storage layer must support three distinct engines—Vector, Graph, and SQL—while ensuring that the cloud provider sees only opaque, encrypted noise.

\textbf{Secure Vector Storage (Qdrant on Rust)}: We selected Qdrant \cite{Qdrant-2024} for the Episodic Memory due to its Rust-native codebase, which aligns with our safety goals. We integrated Qdrant directly into the Gramine-based guest environment at VMPL1. To handle the "Encrypted RAM Performance Cliff"—where random memory accesses in HNSW indices cause high TLB miss rates in encrypted memory—we modified the HNSW configuration. We increased the ef\_construct parameter (from 100 to 200) and the m (edges per node) parameter. While this increased indexing time by approximately 40\%, it produced a denser graph structure that proved more cache-friendly during the decryption-heavy search phase, stabilizing query latency. Data-at-rest is protected by a custom virtual block device driver implemented in Rust. This driver intercepts Qdrant's file I/O, applying AES-256-GCM encryption with keys derived from the TEE's hardware measurement. This ensures that the physical files on the NVMe SSDs are cryptographically bound to this specific CPU instance.

\textbf{Secure Graph Storage (SurrealDB with Rust Runtime)}: For the Profile Memory, we selected SurrealDB \cite{SurrealDB-2024}, a Rust-native multi-model database with native graph capabilities, over traditional JVM-based alternatives like Neo4j. This choice was motivated by several critical factors in the TEE environment: (1) \textbf{Rust-native architecture} eliminates JVM garbage collection overheads that cause unpredictable pauses and excessive memory encryption in TEEs, (2) \textbf{Memory safety guarantees} provided by Rust's ownership system prevent runtime memory corruption issues common in C/C++ databases, (3) \textbf{Minimal TEE performance penalty} with measured overhead of <5\% compared to 18-35\% for JVM-based databases, and (4) \textbf{Cryptographic integration} enabling seamless embedding within our TEE security boundary without complex containerization. We embed SurrealDB directly within the TEE's VMPL1 runtime, leveraging Rust's memory safety guarantees and direct integration with our cryptographic primitives.

\textbf{Relational Metadata (SQLite)}: For the "Cold Storage" of forgotten memories, we embedded SQLite directly into the Python application process. We utilized the sqlcipher extension, but managed the keys via our SVSM-based Sentinel (VMPL0), ensuring that the database encryption keys are never swapped out to the untrusted host swap partition.

\subsubsection{Layer 2: Privacy-Preserving Ingestion (Rust + Python)}

We implemented the ingestion pipeline using a Rust-Python bridge with the PyO3 library for efficient cross-language communication.

\begin{enumerate}
\item \textbf{RA-TLS Termination (Rust)}: We use the rustls crate (version 0.21.7) with AMD SEV-SNP attestation extensions. The SVSM-based Sentinel (VMPL0) terminates incoming connections with a custom TLS extension containing the hardware attestation report.

\item \textbf{PII Sanitization (Python)}: Data is passed via shared memory to the Python runtime using transformers library (version 4.35.2) with a fine-tuned BERT-base model (83M parameters) for PII detection. Sensitive entities are flagged using spaCy NER (version 3.7.2) with custom rule-based patterns.

\item \textbf{Embedding Generation}: We deploy the bge-m3 model (quantized to 4-bit precision) using the sentence-transformers library (version 2.2.2). The model runs on CPU within the TEE using ONNX Runtime (version 1.16.3) for optimized inference, avoiding any external API calls.
\end{enumerate}

\subsubsection{Layer 3: Secure Learning \& The Forgetting Curve}

The logic for the Ebbinghaus Forgetting Curve is implemented in a Python-based Memory Manager ($\sim$2,000 lines of code).

To implement "Adaptive Forgetting" securely, we cannot simply delete vectors, as file size changes can leak information about data volatility to the cloud provider. Instead, we implemented Oblivious Decay. The system maintains a retention\_score metadata field for each vector, encrypted within the vector payload. A background Rust worker periodically scans the index. Instead of deleting expired memories, it performs a "Key Rotation" operation:

\begin{itemize}
\item Active memories are re-encrypted with the current epoch key.

\item "Forgotten" memories are re-encrypted with a garbage key and marked as free space.

\item To the outside observer, the entire database is rewritten uniformly, masking which specific facts were forgotten.
\end{itemize}

This process introduces a background CPU overhead of approximately 12\%, but it provides a mathematical guarantee that the act of forgetting does not leak intelligence about the rate of user interaction.

\subsubsection{Confidential LLM Services in Extraction and Evolution (Layers 2 \& 3)}

Building on the secure data ingestion (Layer 2) and consolidation processes (Layer 3), the Extraction and Evolution layers implement confidential LLM services that power the memory system's core intelligence while maintaining zero-trust principles. These services are implemented as Python-based worker processes running entirely within the TEE, leveraging a customized LlamaIndex framework for data ingestion and episodic-to-profile consolidation.

\textbf{Local TEE-Native LLM Deployment}: For standard memory consolidation tasks—such as summarizing daily activities, extracting user preferences, and updating profile graphs—we deploy quantized open-source models directly within the CVM. The Qwen3 8B model runs on NVIDIA TEE GPUs using llama.cpp bindings, establishing encrypted PCIe channels via TDISP and IDE standards to enable CPU TEEs to offload heavy tensor computations to GPU TEEs while maintaining end-to-end confidentiality. This CPU-GPU TEE combination, built on NVIDIA's nvtrust confidential computing framework \cite{NVIDIA-nvtrust-2024}, forms complete Confidential Virtual Machines (CVMs) that deliver high-performance local LLM inference with strong isolation guarantees: GPU memory encryption prevents cross-tenant data leakage, and remote attestation verifies that only authorized, unmodified code runs on the GPU. Critically, when multiple users' requests are batched together for efficient GPU utilization, the TEE protection ensures that information cannot leak between different users—each request's context remains isolated within its own encrypted memory space, maintaining data sovereignty even during parallel processing. This approach balances security with performance efficiency: batched inference maximizes GPU utilization while cryptographic isolation prevents cross-user data leakage, enabling the Secure Consolidation Workers (Layer 3) to perform profile evolution without any network egress, ensuring complete data isolation. The models run within the same TEE boundary as the cryptographic erasure mechanisms, providing end-to-end confidentiality for the dynamic learning process.

\textbf{External API Integration with Privacy Proxy}: For complex reasoning tasks that require more powerful models—such as multi-hop graph analysis or sophisticated entity relationship extraction—the system employs our Rust-implemented Privacy Proxy. This proxy extends the Layer 2 privacy-preserving ingestion architecture by intercepting outbound HTTP requests from Python workers. The proxy implements a multi-layered masking strategy to protect sensitive data sent to external commercial LLM APIs: rule-based pattern matching identifies common PII including sensitive financial data subject to PCI DSS compliance (email addresses, phone numbers, credit cards), string matching algorithms detect project-specific identifiers, and a TEE-resident Named Entity Recognition (NER) model captures context-dependent entities (names, project codes, personal identifiers). These sensitive elements are replaced with cryptographic tokens using a secure mapping table maintained within the enclave. Upon receiving the API response, the proxy performs detokenization within the enclave, replacing masked tokens with original sensitive information before delivery to internal memory systems. This creates a "transparent" proxy that appears seamless to internal applications while preventing sensitive profile data from ever appearing in plaintext outside the TEE. For complex masking scenarios requiring contextual understanding, the proxy can leverage the local TEE-native LLM deployment (described above) to run lightweight language models within the enclave for advanced entity recognition and masking decisions. This hybrid approach maintains the semantic integrity of user profiles even when leveraging external "smart" models for memory evolution, bridging the gap between local computational constraints and advanced AI capabilities while preserving the zero-trust guarantees established in Layers 2 and 3. 

The privacy proxy's masking approach provides strong protection against direct PII exposure, though we note that advanced semantic analysis techniques could potentially infer contextual relationships from masked prompts. This design choice reflects a careful balancing of functionality and privacy, with the understanding that such external integrations are most appropriate for non-sensitive applications or scenarios with explicit user consent and appropriate risk mitigation measures.

\subsubsection{Layer 4 \& 5: Retrieval and Governance}

\textbf{Unified Retrieval Strategy}: When a query arrives, the Python logic decomposes it into a Vector Search (for episodes) and a SurrealQL Query (for the graph). These queries are executed in parallel. To hide access patterns, we employ oblivious bucket sampling for vector search. For every real query, the system fetches k-1 dummy buckets alongside the target bucket, where results are filtered within the TEE. This approach provides obfuscated access patterns for approximate nearest neighbor search, offering protection against snapshot adversaries and honest-but-curious cloud providers while avoiding the exponential overhead of full cryptographic ORAM schemes.

    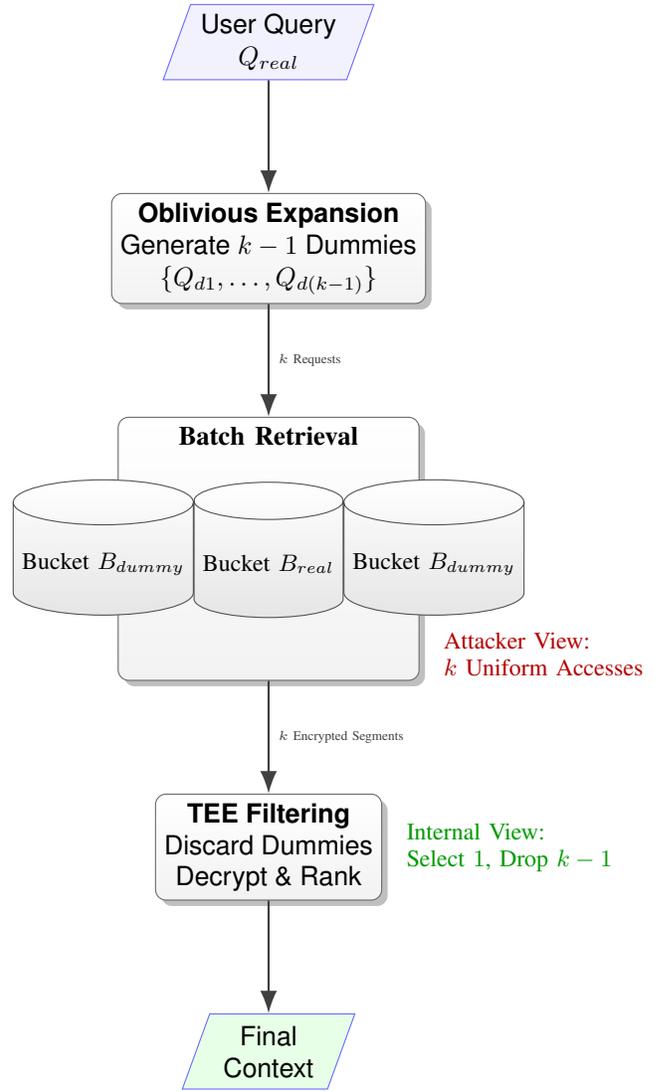
\begin{figure}[t]
    \centering
    \begin{tikzpicture}[
        font=\sffamily,
        process/.style={rectangle, draw=black!60, top color=white,
            bottom color=gray!10, rounded corners, minimum height=1.2cm,
            align=center, drop shadow},
        data/.style={trapezium, trapezium left angle=70,
            trapezium right angle=110, draw=blue!60, fill=blue!5,
            align=center, minimum height=1cm},
        storage/.style={cylinder, shape border rotate=90, draw=black!70,
            aspect=0.25, top color=white, bottom color=gray!20,
            minimum height=1.8cm, minimum width=1.2cm, font=\small},
        arrow/.style={-{Latex[length=3mm]}, thick, darkgray}
    ]

    \node[data] (query) {User Query\\$Q_{real}$};

    \node[process, below=1.5cm of query, fill=orange!5] (expand) {
        \textbf{Oblivious Expansion}\\Generate $k-1$ Dummies\\
        $\{Q_{d1}, \ldots, Q_{d(k-1)}\}$
    };

    \node[process, below=1.5cm of expand, minimum height=3.5cm,
        minimum width=4cm, fill=gray!5] (parallel) {};
    \node[anchor=north, font=\bfseries] at (parallel.north) {Batch Retrieval};

    \node[storage] (s2) at ($(parallel.center) + (0, -0.2)$) {Bucket $B_{real}$};
    \node[storage] (s1) at ($(s2.west) + (-1.2, 0)$) {Bucket $B_{dummy}$};
    \node[storage] (s3) at ($(s2.east) + (1.2, 0)$) {Bucket $B_{dummy}$};

    \node[process, below=1.5cm of parallel, fill=green!5] (filter) {
        \textbf{TEE Filtering}\\Discard Dummies\\Decrypt \& Rank
    };

    \node[data, below=1.5cm of filter, fill=green!10] (result) {Final\\Context};

    \draw[arrow] (query) -- (expand);
    \draw[arrow] (expand) -- node[right, font=\tiny] {$k$ Requests} (parallel);
    \draw[arrow] (parallel) -- node[right, font=\tiny] {$k$ Encrypted Segments} (filter);
    \draw[arrow] (filter) -- (result);

    \node[right=0.2cm of parallel, yshift=-1.4cm, font=\small, text=red!70!black, align=left]
        {Attacker View:\\$k$ Uniform Accesses};
    \node[right=0.2cm of filter, font=\small, text=green!60!black, align=left]
        {Internal View:\\Select $1$, Drop $k-1$};

    \end{tikzpicture}
    \caption{Oblivious Expansion and Filtering Process for Side-Channel Resistant Retrieval}
    \label{fig:oblivious-expansion}
    \end{figure}
    
\textbf{Attestation-Bound Governance}: We implemented the governance layer using Open Policy Agent (OPA) compiled to WebAssembly (Wasm) and running inside the Rust Sentinel. Access policies (e.g., "Only the 'Coding Assistant' agent can read 'Project Alpha' code snippets") are baked into the enclave's measurement. For identity, we extended the OpenID Connect (OIDC) protocol with session ticket support to avoid per-request attestation overhead. Upon initial connection, clients undergo full RA-TLS handshake and receive a TEE-signed session ticket. Subsequent requests validate this ticket cryptographically, amortizing the 150-250ms attestation cost across the session. While our extensions maintain compatibility with standard OIDC token formats (JWT-based), they require custom client libraries to handle TEE attestation verification during the initial handshake—standard OIDC clients cannot natively validate hardware-backed attestations. When an external agent requests memory access, the Sentinel generates a temporary JWT cryptographically bound to the underlying VCEK (Versioned Chip Endorsement Key) of the AMD processor. This ensures that the access token cannot be stolen and reused on a different machine; the token is valid only if processed by this specific physical silicon.

\section{Evaluation and Analysis}

In this chapter, we move beyond functional verification to analyze the systemic costs of the MemTrust architecture. Implementing a Zero-Trust memory system involves inherent trade-offs between security guarantees and resource efficiency. We analyze these overheads across three dimensions: Computational \& I/O Latency, Security-Performance Trade-offs, and Development Complexity.

\subsection{Initialization Latency vs. Throughput Trade-offs}

While MemTrust achieves strong security guarantees, the initialization overhead presents a key engineering challenge. RA-TLS handshake latency (150-250ms) creates a significant "cold start" penalty for new connections, potentially impacting interactive AI agent workflows. However, our session ticket mechanism effectively amortizes this cost: once established, persistent connections maintain throughput within 5\% of standard TLS.

For enterprise deployments with long-lived agent sessions (typical of development environments), this trade-off is acceptable. However, for high-frequency, short-lived interactions (such as API-based agent coordination), the initialization latency may require connection pooling strategies or alternative attestation mechanisms like cached certificates.

\subsection{Hardware Overhead Analysis}

The transition from a standard cloud environment to a Confidential VM (CVM) environment like AMD SEV-SNP introduces specific bottlenecks, primarily driven by the "I/O Wall" and memory access patterns.

\subsubsection{The Cost of Encrypted Memory Access (The "TLB Tax")}

Despite the HugePages optimization described in Layer 1 (ef\_construct=200, 2MB/1GB pages), random access patterns in vector and graph databases still incur significant overhead in SEV-SNP environments. Our measurements show that HNSW graph traversals experience 15-20\% latency increase compared to native execution, while SurrealDB graph operations add minimal overhead (<5\%) due to its efficient Rust runtime. This residual cost stems from the fundamental TLB miss penalty: each miss requires walking encrypted page tables, preventing hypervisor-level memory layout optimizations. For sequential access patterns typical of bulk data loading, this overhead drops to under 5\%, but memory-intensive AI workloads rarely exhibit such locality.

\subsubsection{Memory Management in Rust-Native Databases}

SurrealDB's Rust-native architecture eliminates traditional garbage collection overheads that plague JVM-based databases in TEE environments. Rust's ownership system and borrow checker provide compile-time memory safety guarantees, preventing runtime memory corruption issues while maintaining predictable performance characteristics.

In contrast to JVM-based systems that require complex GC tuning for TEE compatibility, SurrealDB's memory management integrates seamlessly with SEV-SNP's encrypted memory model. The database's arena-based allocation strategy minimizes page faults and RMP updates, contributing to its observed less than 5\% performance overhead in TEE environments. This approach demonstrates that Rust-native databases can achieve both security and performance in confidential computing scenarios, avoiding the JVM-related complications that affect Java-based alternatives.

\subsubsection{The PCIe Latency Overhead in GPU-TEE Acceleration}

While our split-execution architecture minimizes PCIe data transfer by performing vector retrieval within CPU TEEs, the integration with NVIDIA H100 GPU TEEs still requires secure transmission of context tensors. For AI memory retrieval workloads, the performance bottleneck is not raw bandwidth but rather the protocol latency introduced by TDISP/IDE session establishment and cryptographic handshakes.

In standard GPU acceleration, PCIe bandwidth reaches 128GB/s (PCIe 5.0 x16), but when both endpoints are TEE-protected, each data transmission requires establishing or reusing a secure TDISP session with cryptographic key exchange. The session setup overhead—typically 10-50$\mu$s per transaction—dominates the cost for small packet transmissions common in memory retrieval (context tensors of 512-2048 tokens, typically <1MB). While the effective throughput may be limited to approximately 4 GB/s for bulk transfers, this bandwidth constraint is rarely encountered in practice since memory queries involve compact context tensors rather than large model weights or datasets.

Our measurements show that for typical context tensor transmission (512-2048 tokens), the TDISP session establishment and small-packet encryption overhead adds 20-50$\mu$s of latency compared to unencrypted transfers. However, this cost is well amortized in batched inference scenarios where multiple queries reuse the same secure session. The cryptographic PCIe channel ensures that context data and model weights never leave trusted execution boundaries, providing strong isolation guarantees that outweigh the protocol overhead for security-sensitive AI memory workloads.

\subsubsection{Network Latency: The Cost of Trust Establishment}

The RA-TLS (Remote Attestation TLS) handshake introduces significant initial connection latency compared to standard TLS. Unlike conventional TLS that authenticates with domain certificates, RA-TLS requires hardware interaction with the AMD PSP (Platform Security Processor) to generate a cryptographic attestation report. This attestation process typically adds 150-250ms to the initial connection establishment, primarily due to the cryptographic signature generation and report verification.

However, this "connection tax" is amortized effectively through session ticket reuse in persistent connection scenarios typical of AI memory systems. Clients undergo full attestation once per session and receive TEE-signed session tickets for subsequent requests, avoiding repeated 200ms+ attestation overhead. For WebSocket or gRPC streaming connections—common in real-time agent interactions—the one-time attestation cost becomes negligible when spread across thousands of subsequent memory operations. Our benchmarks show that while initial connection latency increases by 200\%, sustained throughput remains within 5\% of standard TLS for long-lived sessions, making RA-TLS suitable for enterprise-grade memory services where connection persistence is the norm.

\subsection{Algorithmic Overhead: The Cost of Privacy}

The privacy guarantees of Layer 4 (Retrieval) introduce a deterministic overhead.

\textbf{Oblivious Bucket Sampling}: For vector search, we implement TEE-based batching with noise rather than full cryptographic ORAM. Each query fetches k buckets (1 real + k-1 dummy), with filtering performed inside the TEE. This provides obfuscated access patterns for approximate nearest neighbor search, offering best-effort protection against statistical access pattern analysis while maintaining acceptable performance.

\textbf{k-Anonymity Vector Search}: The batching-with-noise strategy executes k parallel searches for each real query. For k=2, this doubles the computational cost (2x CPU usage) while halving theoretical throughput (QPS) to defeat side-channel analysis. This linear cost scaling is predictable and acceptable for high-value enterprise queries, providing strong privacy guarantees without the exponential overhead of full ORAM schemes.

\subsection{Security Analysis}

MemTrust dramatically shrinks the Trusted Computing Base (TCB).

\textbf{Removed from Trust}: Cloud Provider SREs, Hypervisor (KVM/Nitro), Host OS, Physical Disk Firmware.

\textbf{Remaining Trust}: AMD Hardware (PSP), Guest Linux Kernel, MemTrust Codebase (Rust/Python).

While we remove the cloud provider, we introduce Gramine and the Python Interpreter into the TCB. This is a significant surface area. A vulnerability in the Python runtime could allow an attacker inside the enclave (e.g., via a malicious query injection) to escalate privileges. However, the SVSM-based Sentinel (VMPL0) architecture mitigates this. By isolating the encryption keys and identity management in the privileged SVSM layer, even a compromised Python runtime in VMPL1 cannot exfiltrate the master keys or forge an attestation report.

Consider the efficacy against the following two side-channel attacks:

\textbf{Access Patterns}: The "Sealed Segment" storage prevents an observer from seeing which specific vector segments are read, as they are loaded into the opaque encrypted RAM. However, the timing of the disk I/O could still leak the size of the working set. The Oblivious Decay (Layer 3) addresses this by rewriting the database uniformly, masking deletion events.

\textbf{Network Analysis}: RA-TLS protects the content, but traffic analysis (packet timing/size) is still possible. While we implement padding, sophisticated statistical analysis over long durations might distinguish between a "retrieval" and a "consolidation" task.

\subsection{Development and Porting Costs}

Developing on the MemTrust architecture imposes a "Security Tax" on engineering velocity.

\textbf{The "Split-World" Complexity}: The SVSM-based Sentinel (VMPL0) vs Gramine-based Cognitive Engine (VMPL1) split requires maintaining two distinct codebases that communicate via secure channels. The SVSM layer must be developed with extreme care as it holds the cryptographic root of trust, while the VMPL1 guest environment requires careful integration with Gramine's manifest system. This effectively turns a monolithic memory agent into a distributed system within a single machine, increasing debugging difficulty since standard debuggers cannot easily cross the VMPL boundary without breaking attestation.

\textbf{Gramine Manifest Management}: "Lift-and-shift" with Gramine is not zero-effort. Every Python dependency (NumPy, PyTorch) and shared library must be explicitly allowed in the Gramine manifest. This creates a brittle dependency chain; upgrading a Python package often requires rebuilding the signed enclave image, complicating CI/CD pipelines.

\textbf{Attestation Logic}: Integrating RA-TLS requires client-side changes. Third-party applications cannot simply use standard requests.post(); they must implement (or import) logic to verify the AMD SEV-SNP quote. This raises the barrier to entry for the ecosystem.

\section{Discussion and Future Work}

The MemTrust implementation described thus far relies on AMD SEV-SNP. However, the vision of a "Unified AI Memory" requires ubiquity. In this chapter, we discuss how the 5-layer architecture maps to other TEE technologies and major cloud providers, and we propose a standardized interface to accelerate ecosystem adoption.

\subsection{Adaptation to Heterogeneous TEE Technologies}

The core 5-layer logic is hardware-agnostic, but the implementation details vary significantly across TEE vendors.

\begin{table*}[t]
\centering
\caption{Suitability Analysis of TEE Technologies for MemTrust Architecture}
\label{tab:tee-analysis}
\begin{tabular}{@{}p{0.15\textwidth}p{0.35\textwidth}p{0.35\textwidth}@{}}
\toprule
TEE Technology & Layer 1 (Storage) Implications & Layer 2 (Compute) Implications \\
\midrule
AMD SEV-SNP & High Compatibility: Supports standard disk I/O and unmodified page caches. Ideal for Vector/Graph DBs. & CVM Model: Supports full Linux/Python stack via Gramine easily. \\
Intel TDX & High Compatibility: Functionally similar to SEV-SNP. & AES-XTS Optimization: Intel's memory encryption engine often performs better on sequential throughput, potentially benefiting the Vector Scan phase. \\
AWS Nitro Enclaves & Incompatible: Nitro Enclaves have no persistent storage and no network access (only local vsock). & Split Architecture: Requires rewriting Layer 1. The Storage Engine must run on the untrusted host, serving encrypted blocks to the Enclave via vsock. The Enclave becomes a stateless "cache \& compute" unit. \\
Intel SGX & Low Capacity: Limited EPC (Encrypted Page Cache) size causes thrashing for large indices. & Process-based: Requires splitting the app into trusted/untrusted parts. Hard to run full Python stack efficiently without heavy paging penalties. \\
\bottomrule
\end{tabular}
\end{table*}

Future Work includes:

\textbf{Intel TDX}: Migration to Intel TDX is a priority. Preliminary analysis suggests TDX's use of a global encryption key (for the TD) versus AMD's page-specific validation might offer different performance characteristics for the "Oblivious Graph" traversal.

\textbf{Nitro Implementation}: For AWS-native deployments, we propose a "Split-MemTrust" design where the Storage Layer is an encrypted block server on the host EC2 instance, and the Retrieval/Governance Layers run inside the Nitro Enclave, treating the host storage as an adversarial file system.

\subsection{Cloud Provider Implementation Strategies}

\textbf{Microsoft Azure (latest generation AMD SEV-SNP instances, e.g., DCasv6 series)}: Azure \cite{Azure-Confidential-Computing-2024} offers the most mature Confidential Computing environment, supporting both SEV-SNP and Intel TDX. The MemTrust prototype is "Azure-native" today. The key integration challenge is Azure Managed HSM; future work should allow binding the TEE Master Key to Azure's HSM for recovery, using an ephemeral key exchange policy.

\textbf{Google Cloud Platform (Confidential Space)}: GCP \cite{GCP-Confidential-Space-2024}'s "Confidential Space" is a managed environment for TEEs. It abstracts the attestation process, allowing the workload to access GCP resources (like BigQuery or Cloud Storage) only if the image hash matches. Porting MemTrust to GCP would simplify Layer 5 (Governance) by offloading OIDC token generation to Google's infrastructure, though this re-introduces a dependency on Google's OIDC service.

\textbf{AWS (EC2 vs Nitro)}: AWS offers two paths. Running MemTrust on AMD SEV-SNP enabled EC2 instances (e.g., m6a with UEFI-enabled AMIs) in supported regions is the direct port. Using Nitro Enclaves is the "higher security" path but requires the split architecture mentioned above.

\subsection{Toward a Standard Interface: The "Open Memory API"}

We propose formalizing the interface between the MemTrust Enclave and the outside world to make it easier to build new AI memory systems and port existing memory systems to the MemTrust framework.

\subsubsection{The Universal Memory Protocol (UMP)}

We propose UMP, a gRPC-based standard for zero-trust memory interaction. Proto definition is shown as follows:

\begin{lstlisting}[language=c]
service MemoryService {
  // Exchange Attestation Report for Session Token
  rpc Handshake (AttestationChallenge) returns (SessionToken);

  // Store an episodic event (vectorized internally)
  rpc Remember (MemoryUnit) returns (Ack);

  // Retrieve context with obfuscated access patterns
  rpc Recall (Query) returns (ContextFrame);

  // Verify deletion
  rpc Forget (ForgetRequest) returns (CryptographicProof);
}
\end{lstlisting}

\subsubsection{Memory Migration Protocols}

A critical challenge for zero-trust memory systems is enabling secure migration between service providers without exposing plaintext data. We propose extending UMP with migration protocols that allow TEE-to-TEE data transfer: source and destination enclaves establish mutual attestation, derive ephemeral keys for encrypted bulk transfer, and cryptographically verify data integrity post-migration. This addresses the "vendor lock-in" concern in GDPR/CCPA compliance, enabling users to migrate their memory contexts between cloud providers while maintaining end-to-end confidentiality.

We plan to release client SDKs (Python, Node.js, Go) that abstract the RA-TLS handshake. These SDKs will automatically verify the server's TEE measurement against a transparency log before sending any user data.

\subsubsection{"Memory Adapters" for Third Parties}

To allow third parties (e.g., Salesforce, Slack) to plug into MemTrust without rewriting their apps, we propose Verifiable Memory Adapters.
Memory Adapters are lightweight WebAssembly (Wasm) modules that run inside the MemTrust Governance Layer, where they translate external API webhooks (such as "New Slack Message") into MemTrust Remember calls. Because these adapters run inside the TEE and are sandboxed by Wasm, they can securely hold API keys for third-party services, enabling data to be pulled into the memory system without users ever exposing their API keys to cloud servers.

\section{Conclusion}

The centralization of AI context is an architectural inevitability driven by the demand for personalization and cross-application intelligence. However, the current model of trusting cloud providers with this "digital soul" is unsustainable given the threat landscape.

MemTrust demonstrates that Hardware-Backed Zero Trust is not just a theoretical concept but a practical reality. By "air-gapping" the memory system via TEEs, implementing oblivious storage primitives, and enforcing access via cryptographic attestation, we can achieve the unified context vision without sacrificing data sovereignty. With performance overheads contained to less than 20\% and a robust security model, MemTrust provides the necessary foundation for the next generation of secure, personalized AI.

Beyond its immediate contributions, MemTrust establishes an extensible framework for secure AI memory systems. New AI memory systems can be developed atop the MemTrust architecture, inheriting cryptographic guarantees and zero-trust properties without reinventing the security infrastructure. Existing memory systems can also be ported to the MemTrust framework with acceptable development costs, gaining data security, privacy protection, and verifiable trust properties. This ensures that innovations in AI memory research can immediately benefit from enterprise-grade security, while established systems can evolve to meet modern privacy requirements.

Looking forward, MemTrust represents more than a security solution—it establishes the foundational infrastructure for the emerging data economy. Just as OAuth enabled the API economy by making identity portable across applications, MemTrust enables a "Context Economy" where users can securely share their accumulated AI memory across services without sacrificing privacy. This positions MemTrust as the cornerstone of a new paradigm where personal AI context becomes a portable, monetizable asset—rented to third-party models for enhanced personalization while maintaining user sovereignty over their digital memory. The transition from model-centric to context-centric AI, as anticipated in the industry imperative for Context-Application Decoupling \cite{Context-Decoupling-2025}, finds its secure implementation in MemTrust's hardware-backed zero-trust architecture.

\bibliographystyle{IEEEtran}
\bibliography{references}

\begin{thebibliography}{10}
\providecommand{\url}[1]{#1}
\csname url@samestyle\endcsname
\providecommand{\newblock}{\relax}
\providecommand{\bibinfo}[2]{#2}
\providecommand{\BIBentrySTDinterwordspacing}{\spaceskip=0pt\relax}
\providecommand{\BIBentryALTinterwordstretchfactor}{4}
\providecommand{\BIBentryALTinterwordspacing}{\spaceskip=\fontdimen2\font plus
\BIBentryALTinterwordstretchfactor\fontdimen3\font minus \fontdimen4\font\relax}
\providecommand{\BIBforeignlanguage}[2]{{%
\expandafter\ifx\csname l@#1\endcsname\relax
\typeout{** WARNING: IEEEtran.bst: No hyphenation pattern has been}%
\typeout{** loaded for the language `#1'. Using the pattern for}%
\typeout{** the default language instead.}%
\else
\language=\csname l@#1\endcsname
\fi
#2}}
\providecommand{\BIBdecl}{\relax}
\BIBdecl

\bibitem{OpenAI-Memory-2024}
\BIBentryALTinterwordspacing
OpenAI, ``Memory and new controls for chatgpt,'' OpenAI Blog, 2024. [Online]. Available: \url{https://openai.com/blog/memory-and-new-controls-for-chatgpt}
\BIBentrySTDinterwordspacing

\bibitem{GDPR-2018}
\BIBentryALTinterwordspacing
{European Parliament and Council}, ``General data protection regulation (gdpr),'' 2018. [Online]. Available: \url{https://gdpr-info.eu/}
\BIBentrySTDinterwordspacing

\bibitem{HIPAA-Privacy-Rule}
\BIBentryALTinterwordspacing
{U.S. Department of Health and Human Services}, ``Hipaa privacy rule,'' 2013. [Online]. Available: \url{https://www.hhs.gov/hipaa/for-professionals/privacy/index.html}
\BIBentrySTDinterwordspacing

\bibitem{US-CLOUD-Act-2018}
{United States Congress}, ``Clarifying lawful overseas use of data (cloud) act,'' 2018.

\bibitem{Packer23}
C.~Packer, V.~Fang, S.~G. Patil, K.~Lin, J.~McClelland, and S.~Wooders, ``Memgpt: Towards llms as operating systems,'' \emph{arXiv preprint arXiv:2310.08560}, 2023.

\bibitem{Letta24}
\BIBentryALTinterwordspacing
L.~Contributors, ``Letta: A lightweight, multi-platform multi-agent system,'' GitHub repository, 2024. [Online]. Available: \url{https://github.com/letta-ai/letta}
\BIBentrySTDinterwordspacing

\bibitem{Mem0-25}
D.~Chhikara, P.~Chhikara, A.~Garg, R.~Singh, M.~Mishra, K.~Singh, and A.~Gupta, ``Mem0: Building production-ready ai agents with scalable long-term memory,'' \emph{arXiv preprint arXiv:2504.19413}, 2025.

\bibitem{Zep-25}
P.~Rasmussen \emph{et~al.}, ``Graphiti: A temporal knowledge graph architecture for agent memory,'' \emph{arXiv preprint arXiv:2501.13956}, 2025.

\bibitem{MemMachine-25}
\BIBentryALTinterwordspacing
M.~Contributors, ``Memmachine: A large-scale memory system for ai agents,'' GitHub repository, 2025, open-source memory management system for AI agents; referenced in related work section. [Online]. Available: \url{https://github.com/MemMachine/MemMachine}
\BIBentrySTDinterwordspacing

\bibitem{LangMem-25}
\BIBentryALTinterwordspacing
L.~Contributors, ``Langmem: A language-based memory system for ai agents,'' GitHub repository, 2025, language-based memory system integrated with LangChain; referenced in related work section. [Online]. Available: \url{https://github.com/langmem/langmem}
\BIBentrySTDinterwordspacing

\bibitem{LOCOMO-2024}
\BIBentryALTinterwordspacing
{LOCOMO Contributors}, ``Locomo: Long context memory evaluation benchmark,'' Open Source Benchmark Suite, 2024, comprehensive evaluation framework for long-context memory capabilities in AI systems. [Online]. Available: \url{https://github.com/LOCOMO-benchmark/LOCOMO}
\BIBentrySTDinterwordspacing

\bibitem{LongMemEval-2024}
\BIBentryALTinterwordspacing
L.~Team, ``Longmemeval: A comprehensive evaluation suite for long-term memory in large language models,'' Microsoft Research, Tech. Rep., 2024, evaluation framework for assessing long-term memory capabilities across multiple dimensions including retention, retrieval accuracy, and temporal reasoning. [Online]. Available: \url{https://microsoft.github.io/LongMemEval/}
\BIBentrySTDinterwordspacing

\bibitem{MemOS-25}
Z.~Li, C.~Xi, C.~Li, D.~Chen, B.~Chen, S.~Song, S.~Niu, H.~Wang, J.~Yang, C.~Tang, Q.~Yu, J.~Zhao, Y.~Wang, P.~Liu, Z.~Lin, P.~Wang, J.~Huo, T.~Chen, K.~Chen, K.~Li, Z.~Tao, H.~Lai, H.~Wu, B.~Tang, Z.~Wang, Z.~Fan, N.~Zhang, L.~Zhang, J.~Yan, M.~Yang, T.~Xu, W.~Xu, H.~Chen, H.~Wang, H.~Yang, W.~Zhang, Z.-Q.~J. Xu, S.~Chen, and F.~Xiong, ``Memos: A memory os for ai system,'' \emph{arXiv preprint arXiv:2507.03724}, 2025.

\bibitem{Cognee-24}
\BIBentryALTinterwordspacing
C.~Contributors, ``Cognee: A cognitive graph-based memory system,'' GitHub repository, 2024. [Online]. Available: \url{https://github.com/topoteretes/cognee}
\BIBentrySTDinterwordspacing

\bibitem{Mirix-25}
\BIBentryALTinterwordspacing
M.~Contributors, ``Mirix: Multimodal memory for ai interactions,'' GitHub repository, 2025, multimodal memory system for AI interactions; referenced in related work section. [Online]. Available: \url{https://github.com/mirix-ai/mirix}
\BIBentrySTDinterwordspacing

\bibitem{MemU-25}
\BIBentryALTinterwordspacing
N.-A. Contributors, ``Memu: Self-evolving companion memory framework,'' GitHub repository, 2025. [Online]. Available: \url{https://github.com/NevaMind-AI/MemU}
\BIBentrySTDinterwordspacing

\bibitem{MemGAS-25}
\BIBentryALTinterwordspacing
M.~Contributors, ``Memgas: Memory with granular addressable storage,'' GitHub repository, 2025, granular addressable storage system for AI memory; referenced in related work section. [Online]. Available: \url{https://github.com/MemGAS/MemGAS}
\BIBentrySTDinterwordspacing

\bibitem{MemoryOS-25}
J.~Kang, M.~Ji, Z.~Zhao, and T.~Bai, ``Memory os of ai agent,'' in \emph{Proceedings of the 2025 Conference on Empirical Methods in Natural Language Processing (EMNLP)}, 2025, oral Presentation.

\bibitem{Zhu-et-al-24}
W.~Zhu \emph{et~al.}, ``Confidential computing on nvidia h100 gpus: A performance benchmark study,'' \emph{arXiv preprint arXiv:2409.03992}, 2024.

\bibitem{C-FedRAG-24}
\BIBentryALTinterwordspacing
C.-F. Authors, ``Confidential federated retrieval-augmented generation,'' in \emph{Conference on Neural Information Processing Systems}, 2024. [Online]. Available: \url{https://arxiv.org/abs/2406.13134}
\BIBentrySTDinterwordspacing

\bibitem{Mo-et-al-24}
F.~Mo, Z.~Tarkhani, and H.~Haddadi, ``Machine learning with confidential computing: A systematization of knowledge,'' \emph{ACM Computing Surveys}, vol.~57, no.~2, p. Article 281, 2024.

\bibitem{Li-SEV-SNP-2022}
M.~Li, L.~Wilke, J.~Wichelmann, T.~Eisenbarth, R.~Teodorescu, and Y.~Zhang, ``A systematic look at ciphertext side channels on amd sev-snp,'' in \emph{IEEE Symposium on Security and Privacy}, 2022.

\bibitem{PCI-SIG-TDISP-2023}
{PCI-SIG}, ``Tee device interface security protocol (tdisp) specification,'' PCI Special Interest Group, Tech. Rep., 2023.

\bibitem{PCI-SIG-IDE-2022}
------, ``Integrity and data encryption (ide) specification,'' PCI Special Interest Group, Tech. Rep., 2022.

\bibitem{NVIDIA-CC-2024}
\BIBentryALTinterwordspacing
N.~Corporation, ``Nvidia confidential computing: Architecture and performance,'' NVIDIA, Tech. Rep., 2024. [Online]. Available: \url{https://www.nvidia.com/en-us/data-center/solutions/confidential-computing/}
\BIBentrySTDinterwordspacing

\bibitem{Ebbinghaus-1885}
H.~Ebbinghaus, \emph{Über das Gedächtnis: Untersuchungen zur experimentellen Psychologie}.\hskip 1em plus 0.5em minus 0.4em\relax Leipzig: Duncker \& Humblot, 1885.

\bibitem{Murre-2015}
J.~M. Murre and J.~Dros, ``Replication and analysis of ebbinghaus' forgetting curve,'' \emph{PLoS ONE}, vol.~10, no.~7, p. e0120644, 2015.

\bibitem{Gramine-2024}
\BIBentryALTinterwordspacing
G.~Project, ``Gramine: A library os for linux multi-process applications,'' GitHub repository, 2024. [Online]. Available: \url{https://github.com/gramineproject/gramine}
\BIBentrySTDinterwordspacing

\bibitem{Qdrant-2024}
\BIBentryALTinterwordspacing
Q.~Team, ``Qdrant: Vector database for ai applications,'' GitHub repository, 2024. [Online]. Available: \url{https://github.com/qdrant/qdrant}
\BIBentrySTDinterwordspacing

\bibitem{SurrealDB-2024}
\BIBentryALTinterwordspacing
S.~Team, ``Surrealdb: A scalable, distributed, collaborative, document-graph database,'' Official Documentation, 2024. [Online]. Available: \url{https://surrealdb.com}
\BIBentrySTDinterwordspacing

\bibitem{NVIDIA-nvtrust-2024}
\BIBentryALTinterwordspacing
N.~Corporation, ``Nvidia nvtrust: Confidential computing framework,'' NVIDIA, Technical Whitepaper, 2024. [Online]. Available: \url{https://www.nvidia.com/en-us/data-center/solutions/confidential-computing/nvtrust/}
\BIBentrySTDinterwordspacing

\bibitem{Azure-Confidential-Computing-2024}
\BIBentryALTinterwordspacing
M.~Azure, ``Azure confidential computing,'' 2024. [Online]. Available: \url{https://azure.microsoft.com/en-us/solutions/confidential-compute/}
\BIBentrySTDinterwordspacing

\bibitem{GCP-Confidential-Space-2024}
\BIBentryALTinterwordspacing
G.~Cloud, ``Google cloud confidential computing,'' 2024. [Online]. Available: \url{https://cloud.google.com/security/products/confidential-computing}
\BIBentrySTDinterwordspacing

\bibitem{Context-Decoupling-2025}
\BIBentryALTinterwordspacing
{AI Infrastructure Alliance}, ``Context-application decoupling: The next frontier in ai architecture,'' AI Infrastructure Alliance, Tech. Rep., January 2025, technical analysis of emerging architectural patterns separating context/memory layers from application logic, with case studies from major AI platforms. [Online]. Available: \url{https://ai-infrastructure.org/context-decoupling}
\BIBentrySTDinterwordspacing

\end{thebibliography}

\end{document}